\documentclass[fleqn,usenatbib]{mnras}

\usepackage{subfiles}
\usepackage{newtxtext,newtxmath}

\usepackage[T1]{fontenc}

\DeclareRobustCommand{\VAN}[3]{#2}
\let\VANthebibliography\thebibliography
\def\thebibliography{\DeclareRobustCommand{\VAN}[3]{##3}\VANthebibliography}

\usepackage{graphicx}	
\usepackage{amsmath}	

\usepackage{amssymb}	
\usepackage{xspace}

\newcommand{\kmsmpc}{\kms\;{\rm Mpc}^{-1}}

\newcommand{\hkpc}{h^{-1}{\rm kpc}}
\newcommand{\hmpc}{h^{-1}{\rm Mpc}}

\newcommand{\mpccm}{\rm cMpc}

\newcommand{\kms}{\;{\rm km}\,{\rm s}^{-1}}

\newcommand{\msolar}{\;{\rm M}_{\odot}}
\newcommand{\mstellar}{\;{\rm M}_{\star}}
\newcommand{\msolaryr}{\;{\rm M}_{\odot} {\rm yr}^{-1}}

\newcommand{\simba}{\mbox{{\sc Simba}}\xspace}
\newcommand{\simbac}{\mbox{{\sc Simba-C}}\xspace}
\newcommand{\illustris}{\mbox{{\sc IllustrisTNG}}\xspace}
\newcommand{\gizmo}{\mbox{{\sc Gizmo}}\xspace}

\newcommand{\grackle}{\mbox{{\sc Grackle-3.1}}\xspace}
\newcommand{\chem}{\mbox{{\sc Chem5}}\xspace}

\newcommand{\fgas}{f_{\rm gas}}
\newcommand{\fbh}{f_{\rm BH}}
\newcommand{\fdust}{f_{\rm dust}}
\newcommand{\fedd}{f_{\rm Edd}}
\newcommand{\mbh}{\;{\rm M}_{\rm BH}}

\newcommand{\caesar}{\mbox{{\sc Caesar}}\xspace}

\title[Early Quenched Galaxies in \simba]{The Nature and Evolution of Early Massive Quenched Galaxies in the \simbac Simulation}
\author[]{Jakub Szpila$^1$, Romeel Dav\'e$^{1,2}$, Douglas Rennehan$^{3}$, Weiguang Cui$^{1,4,5}$, Renier T. Hough$^{6}$
\\
$^{1}$ Institute for Astronomy, University of Edinburgh, Royal Observatory, Blackford Hill, Edinburgh EH9 3HJ, UK\\
$^{2}$University of the Western Cape, Bellville, Cape Town 7535, South Africa \\
$^{3}$Center for Computational Astrophysics, Flatiron Institute, 162 Fifth Avenue, New York, NY, 10010, USA\\
$^{4}$ Departamento de Física Teórica, M-8, Universidad Autónoma de
Madrid, Cantoblanco 28049, Madrid, Spain\\
$^{5}$ Centro de Investigación Avanzada en Física Fundamental,(CIAFF), Universidad Aut\'{o}noma de Madrid, Cantoblanco, 28049 Madrid, Spain\\
$^{6}$Centre for Space Research, North-West University, Potchefstroom 2520, South Africa
}

\date{Accepted XXX. Received YYY; in original form ZZZ}

\pubyear{2024}

\begin{document}
\label{firstpage}
\pagerange{\pageref{firstpage}--\pageref{lastpage}}
\maketitle

\begin{abstract}
We examine the nature, origin, and fate of early ($z\geq 2$) massive ($M_\star>10^{10}M_\odot$) quenched galaxies (EQGs) in a new $(100\hmpc^3)$ run of the \simbac galaxy formation model.  We define ``quenched'' to be $>4\sigma$ below an iterative polynomial fit to the star-forming sequence (SFS), and find that \simbac produces EQGs as early as $z\sim 5$ and number densities agreeing with observations at $z\la 3$ (though slightly low at $z\ga 4$). Using a photometric-based EQG selection or a fixed sSFR cut of $10^{-10}$yr$^{-1}$ yields similar results. EQGs predominantly arise in central galaxies with stellar mass $M_\star\sim 10^{10.5-11.3}M_\odot$, not necessarily the most massive systems. A UMAP projection shows that quenched galaxies have notably large black hole-to-stellar mass ratios, lower rotational support, and less dust, but are not atypical versus similar-mass non-EQGs in their environments, halo mass, or halo gas temperatures at the time of quenching. However, via galaxy tracking we show that the progenitor environments of EQGs are significantly more overdense than that of non-EQGs, which drives higher black hole mass fractions and stellar-to-halo mass ratios.  This results in the Eddington ratio dropping sufficiently low for \simbac's jet mode feedback to turn on, which quickly quenches the host galaxies. EQGs thus seem to be galaxies that grow their black holes quickly within highly dense environments, but end up in moderately-dense environments where black hole feedback can quench effectively.  We find that $\ga 30\%$ of EQGs rejuvenate, but the rejuvenating fraction drops quickly below $z\la 2$.  By $z=0$ it is difficult to distinguish the descendants of EQGs vs. non-EQGs.
\end{abstract}


\section{Introduction} \label{Introduction}
A major result from large multi-wavelength galaxy surveys such as the Sloan Digital Sky Survey (SDSS; \citealt{York2000}) has been the classification of galaxies in the local Universe into two distinct groups by their colour: star-forming (or 'main-sequence') galaxies, which are typically blue with spiral morphology, and quiescent (or 'quenched') galaxies, which are red and elliptical \citep{Blanton2003, Kauffmann2003, Baldry2004}. This classification has been extended to galaxies beyond the local Universe, with evidence of massive quiescent galaxies as far as $z\sim 3$ \citep{Tomczak2014, Man2016, Martis2016} and even $z\sim 4$ \citep{Straatman2015}. The sensitive near-infrared capabilities of the James Webb Space Telescope (JWST) have pushed out the current record-holder to $z\approx 4.658$, just 1.3~Gyr after the Big Bang \citep{Carnall2023}.

Massive quiescent galaxies with stellar mass approaching $M_\star\sim 10^{11}M_\odot$ or more appearing so early in cosmic history is a significant surprise.  Assembling that much mass in of itself is not a challenge within the $\Lambda$ Cold Dark Matter ($\Lambda$CDM) cosmological model~\citep{Behroozi2018}, but it may suggest that star formation-related feedback processes required to self-regulate growth in the bulk of the galaxy population even at very early epochs~\citep[e.g.][]{Dave2006} are for some reason strongly curtailed in some subset of systems~\citep{Dekel2023, Rennehan2024manhattanarxiv}.  What makes these systems special is unclear, since there are many more equally massive galaxies at those epochs that are not quenched.

An even bigger mystery is why these galaxies stop forming stars.  Modern galaxy formation models invoke active galactic nuclei (AGN) feedback to quench galaxies~\citep[e.g.][]{SomervilleDave2015}. Current ideas of galaxy quenching include being a result of a major merger, where strong supernova and black hole feedback lead to a blowout of gas and dust, removing the material needed for star formation \citep{Hopkins2008}. Another avenue is to invoke feedback from AGN in order to heat the gas surrounding the galaxy, which suppresses gas accretion and eventually quenches the galaxy~\citep{Croton2006,Bower2006}. For early quenched systems, one might expect the merger path to be more appropriate, since mergers are more common at early times, blowout allows rapid quenching, and circum-galactic gas is quite dense making halo heating less effective.  However, \citet{Rodriguez2022} showed that at least in one model rapid quenching need not be correlated with major merger activity, and merger-driven blowout alone cannot permanently quench galaxies since accretion would restart rapidly~\citep[e.g.][]{Gabor2011} particularly in the early Universe when accretion rates are high~\citep{Dekel2009}. Most galaxy formation models require some form of halo-based quenching in order to reproduce the quenched galaxy population as seen today~\citep{SomervilleDave2015}, but the situation is less clear for early quenched systems.

Cosmologically-situated hydrodynamic simulations of galaxy formation have emerged as the leading approach to understand the growth and evolution of galaxies within a hierarchical context~\citep{Vogelsberger2020}.  While such models can broadly reproduce the overall growth of the galaxy population, they generally struggle to produce enough early ($z\ga 3$) massive quenched galaxies as are now observed~\citep{DeLucia2024}.  This suggests that there is some missing physics in the models, or at least that the current implementations are not flexible enough to produce the most extreme objects at these early epochs. Such observations therefore offer an opportunity to improve such models, which could have implications for understanding the overall evolution of the massive galaxy population to lower redshifts.

To quantify the evolution of quenched galaxies, one must first define what is quenched, which is already a murky topic.  In the low-redshift Universe, quenched galaxies can be defined as lying along the red sequence in the colour-magnitude diagram~\citep{Blanton2003}. But at intermediate redshifts, contamination by dusty red galaxies favours using two-colour diagrams such as UVJ~\citep{Patel2013}. With improved multi-wavelength surveys combined with sophisticated spectral energy distribution (SED) fitting codes~\citep[e.g.][]{Leja2019,Carnall2019}, it becomes possible to measure $M_\star$ and SFR reasonably accurately out to high redshifts, enabling a definition in terms of the specific star formation rate sSFR$\equiv$SFR/$M_\star$.  Locally, a canonical definition for quenched galaxies is sSFR$<10^{-11}$yr$^{-1}$, which lies about an order of magnitude below star-forming galaxies of similar mass.  But at higher redshifts, it is measured that the galaxy main sequence (i.e. SFR vs. $M_\star$) shows higher SFR at a given $M_\star$ at earlier epochs \citep{Daddi2007, Elbaz2007, Salim2007, Speagle2014,Davies2016}.  Using the same local sSFR cut to define quenched galaxies thus may not be appropriate in the early Universe, and instead it should account for the location of the main sequence at the relevant redshift.

Quenched galaxies may not remain so indefinitely. In models where galaxy accretion is related to halo accretion, following a period of quiescence, a galaxy can experience a secondary period of star formation or ``rejuvenation," when enough gas and dust re-accretes onto the galaxy for it to return to the main sequence.  It is estimated that $\sim 10-30\%$ of galaxies in the local universe undergo such a process \citep{Donas2007, Schawinski2007, Pandya2017, Behroozi2019}, depending on how one defines rejuvenation. It is unclear how this fraction evolves; one might expect the rejuvenation fraction to decrease with increasing redshift, since there has been less time for rejuvenation to occur, but it could also increase to higher redshifts if the more rapid cooling allows accretion to restart more rapidly. 

In this paper, we examine the nature of early quenched galaxies seen in the \simbac~\citep{Hough2023} simulation (described in \S\ref{sec:simba}). \simbac is an update of the widely used \simba simulation~\citep{Dave2019} that updates the chemical enrichment and stellar feedback processes and recalibrates AGN feedback, yielding a more physically motivated model that improves abundance ratios \citep{Hough2023} and X-ray properties \citep{Hough2024}. Both the chemical enrichment and the way how stellar feedback ejects energy and mass were updated to be time-step resolved. Thus it provides a plausible platform to study the evolution of quenched massive galaxies.  

This paper is organised as follows.  To study the massive quenched population, we begin by defining our quenched sample at various redshifts based on the star-forming sequence (SFS).  We then use Uniform Manifold Approximation and Projection (UMAP) (\S\ref{sec:UMAP}) to qualitatively identify the properties that are uniquely shared by early quenched galaxies, and quantify this via comparisons to a stellar mass-matched sample of star-forming galaxies.  We then track galaxy progenitors and descendants to examine how the AGN feedback interplays with the quenching process, and quantify how often early quenched systems end up rejuvenating.  We investigate the energetics of AGN feedback to see how this quenches galaxies, and finally examine the $z=0$ descendants of quenched galaxies to attempt to identify any characteristics that early quenched galaxies share today.  Finally, we present our conclusions in \S\ref{sec:conclusion}. 

\section{The \simbac Simulation} \label{sec:simba}

This analysis focuses on the quenched galaxies present at high redshift in the \simbac simulation. Some details of the implementation of processes relevant for quenching in \simbac are described below, adapted from \citet{Dave2019} and \citet{Hough2023} where a fuller description can be found.  We further describe our approach for identifying quenched galaxies at each redshift based on the star-forming main sequence.

\subsection{\simbac}

We run and analyse a $(100\hmpc)^3$ volume with $2\times 1024^3$ elements evolved with the \simbac model~\citep{Hough2023, Hough2024}. \simbac is an improved version of the \simba model~\citep{Dave2019}, whose primary new component is a state-of-the-art chemical evolution model but also includes tweaks to the feedback parameters and subgrid implementations. \citet{Hough2023} ran a $(50\hmpc)^3$ volume with $2\times 512^3$ elements, and our run replicates many of those results while yielding a much enlarged sample of rare massive galaxies.  Our $100\hmpc$ run \citep{Hough2024} uses the same initial conditions at $z=249$ as the original flagship \simba run, with the same minimum gravitational softening length of $0.5\hkpc$ and the same {\it Planck}-concordant cosmology of $\Omega_m=0.3$, $\Omega_\Lambda=0.7$, $\Omega_b=0.048$, $H_0=68\kmsmpc$, and $n_s=0.97$, where these quantities have their usual cosmological meanings.  We output 152 snapshots from $z=99\to 0$ just as in \simba.  For a complete description of the \simbac model and a detailed analysis of the differences versus \simba, see \citet{Hough2023, Hough2024}.

\simbac, like \simba, uses the hydrodynamics and gravity solver modules from \gizmo \citep{Springel2005, Hopkins2015, Hopkins2017}, evolving the hydrodynamics equations using the meshless finite mass method. Shocks are handled using a Riemann solver with no artificial viscosity, providing better handling of strong shocks and shear flows.  To model radiative cooling and photoionisation heating of gas \simba uses the \grackle library \citep{Smith2017} with metal cooling and non-equilibrium evolution of primordial elements. For accuracy and stability, the adiabatic and radiative terms are evolved together isochorically on the cooling timescale.

Star formation is modelled using an 
$\text{H}_{2}$-based star formation rate (SFR), where the $\text{H}_{2}$ fraction is computed based on the subgrid model of \citet{Krumholz2011} based on the metallicity and column density of $\text{H}_{2}$ with minor modifications to account for numerical resolution as described in \citet{Dave2016}. SFR is given by $\text{H}_{2}$ density scaled by the star formation parameter $\epsilon_{*}=0.026$ \citep{Kennicutt1998, Pokhrel2021} divided by the dynamical time: $\text{SFR}=\epsilon_{*}\rho_{\text{H}_{2}}/\text{t}_{\text{dyn}}$.  Stars are converted from gas particles probabilistically. 

\simbac introduced a new chemical enrichment model called \chem, developed by \citet{Kobayashi2014}, tracking all elements from hydrogen (H) to germanium (Ge).  At the same time, it removes \simba's assumption of instantaneous recycling of metals and energy from Type II supernovae as well as an ad hoc model for Type Ia SNe, replacing it with a full tracking of the metal return from Type II+Ia supernovae and stellar mass loss on the correct stellar evolution timescales.  For a full description see \citet{Hough2023}.

Stellar feedback is handled via a Monte Carlo prescription for kinetic outflows, in which each galaxy is assigned a mass loading factor based on its stellar and gas properties, and then if selected a gas particle will become a wind particle and be kicked with a velocity scaled to the galaxy's escape velocity.  The detailed equations are presented in \citet{Hough2023}, which are slightly updated from those in the original \simba model given in \citet{Dave2019}. These wind particles also experience a brief period of ``decoupling" where their hydrodynamic forces are shut off, in order to mimic channels that allow relatively unfettered escape from the interstellar medium. 

Black holes are seeded and grown during the simulation, with accretion energy used to drive galaxy quenching through feedback. Black holes are seeded when the galaxy is initially resolved $\mstellar \gtrsim 6\times 10^{8}\msolar$ ($\approx 32$ star particle masses) as opposed to the higher threshold in \simba of $\mstellar \gtrsim 5\times 10^{9}\msolar$. To mimic the behaviour of star formation in dwarf galaxies as described in \citet{Angles2017b, Hopkins2022} black hole accretion is exponentially suppressed with a factor of $\text{exp}(-\mbh/10^{6}\msolar)$. This reduces some artificial features in the galaxy population owing to this threshold including a 'pile-up' of galaxies just above the BH seeding threshold seen in \simba \citep{Dave2019}. Moreover, the lower seeding mass contributes to the increased number of early quenching galaxies in \simbac as we will discuss in \S\ref{SFS} and \S\ref{Mean evo}.

Given the importance of AGN feedback, we describe its implementation in \simbac in more detail.  \simbac follows \simba in which AGN feedback is split into two regimes: radiative winds, which are intended to crudely represent molecular and ionised outflows from the black hole with velocities of $\sim10^{3}\ {\rm kms}^{-1}$ \citep{Sturm2011, Greene2012, Liu2013, Perna2017}) and jets reaching velocities of $10^{4}\ {\rm kms}^{-1}$ or more of highly ionised material.  According to \citet{Best2012, Heckman2014}, these modes are associated with different Eddington ratios of the black hole, where the Eddington ratio is the black hole accretion rate in units of the Eddington accretion rate, with low Eddington rates resulting in jets, while high Eddington rates are associated with radiative winds. Specifically, \simba uses bipolar kinetic feedback from AGN, decoupled for a short time ($10^{-4}$ of a Hubble time at launch), with a wind speed given by the following formulae for higher Eddington ratios ($f_{\rm Edd}>0.2$, quasar mode):
\begin{equation} \label{radv}
    v_{w, rad}=500 + 500 \log\left(\frac{\mbh}{10^6 \msolar}\right)^{1/3}\ {\rm kms}^{-1}.
\end{equation}
At lower $f_{\rm Edd}$ ($f_{\rm Edd}<0.2$) there is a transition to low-$f_{\rm Edd}$ mode (jets) becoming stronger as $f_{\rm Edd}$ decreases until it reaches a maximum at $f_{\rm Edd}=0.02$ (jet mode):
\begin{equation} \label{jetv}
    v_{w, jet}=v_{w, rad}+7000\log\left(\frac{0.2}{MAX(f_{\rm Edd}, 0.02)}\right)\ {\rm kms}^{-1}.
\end{equation}
This is where there is a slight change from \simba: \simbac adds a dependence of the jet velocity cap on halo escape velocity from a flat maximum of the jet velocity term of $7000\kms$. Additionally the minimum black hole mass constraint is increased to $7\times10^{7}\msolar$, to avoid jet mode occurring due to high stochasticity in poorly-resolved small galaxies. All outflows have a momentum input based on the accretion rate of the black hole:
\begin{equation} \label{momentum}
    \dot{P}_{out}=20\frac{L}{c},
\end{equation}
where $L=\eta\dot{M}_{\rm BH}c^{2}$ is the bolometric luminosity of the AGN and $\eta=0.1$ is the radiative efficiency.  All outflows are launched in a collimated bipolar fashion, based on the angular momentum of the inner disk from which the black hole is accreting. Additional X-ray feedback is introduced mimicking X-ray heating from the black hole accretion disk in galaxies with $f_{gas}<0.2$ and with full velocity jets, following \citet{Choi2012}.

\subsection{Galaxies, Halos, and Photometry}

Halos are identified during the simulation run via a 3-D friends-of-friends (FoF) algorithm with a linking length set to 0.2 times the mean inter-particle separation.  From within each halo, we identify galaxies using \caesar, which uses a 6-D FoF with a much smaller linking length and a velocity linking length equal to the local velocity dispersion.  Only star particles and gas elements eligible for star formation with $n_H>0.13$cm$^{-3}$ are used in galaxy identification.  The particle lists from the galaxy/halo are then used to compute a wide range of physical properties, the computation of which we will describe when we discuss specific properties.

Using the stars, \caesar produces galaxy photometry using the Flexible Stellar Population Synthesis code \citep[FSPS;][]{Conroy2009, Conroy2010}.  We treat each star particle as a single stellar population (SSP) of a given mass, age, and metallicity, and assuming a \citet{Chabrier2003} IMF, determine the star's spectrum including emission lines.  We then extinct the star's spectrum along a chosen line of sight, using the \simbac-tracked dust along the line of sight.  A galaxy's spectrum is then the sum of the spectra of all the stars in the galaxy, to which chosen bandpasses are applied to obtain magnitudes.  All the above information, including galaxy and halo physical and photometric properties, are stored in a single {\tt hdf5} \caesar catalog file for each simulation snapshot.

\section{Identifying quenched galaxies} \label{SFS}
\begin{figure*}
    \includegraphics[width=0.95\linewidth]{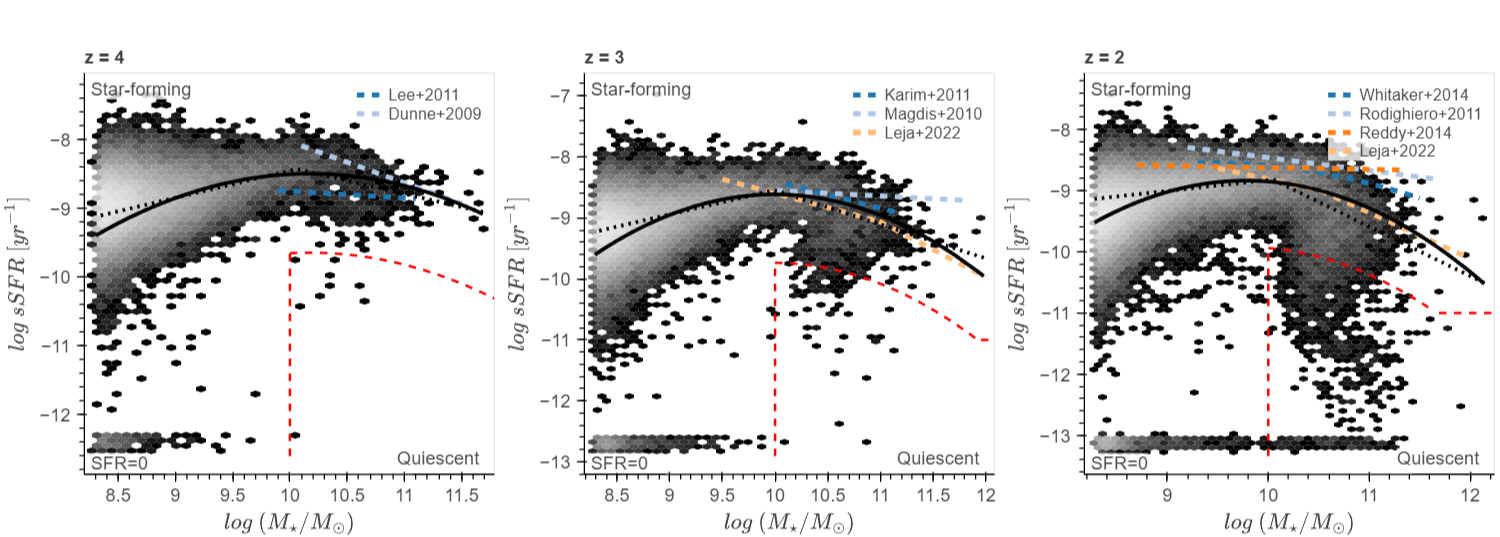}
    \caption{The ${\rm sSFR}-\mstellar$ relation in \simbac at $z=4,3,2$, lighter bins have higher galaxy density. The solid and dotted black lines show respectively the quadratic and broken power law best fits for the main sequence. The red dashed line represents the cut-off for identifying quenched galaxies. Overall we derive a star-forming main sequence in line with literature results at high redshift and with the more recent census at $z = 3,2$, significantly below previous results. Most quenched galaxies are not highly quenched (${\rm sSFR}<10^{-11}yr^{-1})$ but are still significantly below the star forming population. The pile-up at low sSFR is due to a number of galaxies with ${\rm SFR}=0$, scattered around the minimum non-zero SFR.}
    \label{fig:SFSfit}
\end{figure*}

To explore distinctive features and behaviour of quenched galaxies, their population must first be identified within \simbac.  There are numerous definitions of quenched galaxies, but here we focus on two approaches that have been used at high-$z$, one more theoretical-based and another more observationally-motivated.

Theoretically, a quenched galaxy is typically defined as lying below a particular threshold of star formation rate ($SFR$) or specific star formation rate ($sSFR=\frac{SFR}{M_\star}$).  However, using a fixed threshold for sSFR is a poor choice for evolutionary studies since the star-forming sequence evolves significantly \citep{Speagle2014}.  Therefore, a more redshift-agnostic choice is to identify galaxies as quenched if they are lying substantially below the main sequence at the epoch of that galaxy.  This is the definition we choose here.

We fit the $\log {\rm sSFR}-\log \mstellar$ relation individually to each \simbac snapshot with a second order polynomial equation considering at first galaxies with $\mstellar>10^9\msolar$ and ${\rm sSFR} > 10^{-11}{\rm yr}^{-1}$:

\begin{equation}
    \log{\rm sSFR}=a\log{\mstellar}^2+b\log{\mstellar}+c
\end{equation}

To fit the SFS only with star-forming galaxies we then iteratively remove low SFR outliers, i.e. galaxies $>2\sigma$ below the best-fit, where $\sigma$ is the residual standard deviation, and re-fit the relation, until the sum of squared residuals does not decrease. A second-order polynomial fit is chosen to reflect recent findings that the SFS flattens in the high stellar mass regime \citet{Lee2015, Grootes2017, Grootes2018}, and remains relatively flat at high redshift (\citet{Lee2015, Schreiber2015}). At the redshifts we consider, this is a very stable procedure because the large majority of galaxies lie along a well-defined main sequence. The resulting best-fit coefficients and uncertainties (obtained using Scipy's {\sc curve\_fit}) for redshifts $z=2,3,4$ are presented in Table \ref{tab:sfs_fit}.

\begin{table}
    \caption{Polynomial coefficients for quadratic main sequence fits.}
    \label{tab:sfs_fit}
    \begin{tabular}{cccc}
        \hline
        redshift & a & b & c\\
        \hline
        2 & $-0.303\pm0.007$ & $5.92\pm0.13$ & $-37.8\pm0.7$\\
        3 & $-0.340\pm0.008$ & $6.80\pm0.17$ & $-42.6\pm0.8$\\
        4 & $-0.257\pm0.013$ & $5.23\pm0.25$ & $-35.1\pm1.2$\\
        \hline
    \end{tabular}
\end{table}

Quenched galaxies are then chosen as those lying $>4\sigma$ below the SFS best-fit line or ${\rm sSFR}<10^{-11}{\rm yr}^{-1}$. Because we want to study the more massive population that has been observed recently, we further make a cut at $\mstellar>10^{10}\msolar$; typically, lower mass quenched galaxies are smaller satellite systems, which possibly originate due to different quenching mechanisms compared to the massive galaxies. This then defines our quenched galaxy population.

The $\sigma$ thresholds were selected to minimise sum of squared residuals. Variation in the procedure and the final quenched galaxies selection results in detecting roughly 10-20\% more or less quiescents, and does not change the broader conclusions of this work.

The selection of quiescent galaxies at $z=4,3,2$ is shown in Figure \ref{fig:SFSfit}. The solid black line shows the main sequence second-order polynomial fit to the galaxy population, the dotted black line shows the broken power law fit for comparison. Massive quiescent galaxies lie below the red dashed line, determined by the quadratic fit, which at $z=4$ roughly corresponds to sSFR$\la 10^{-10}$yr$^{-1}$ and shows stronger mass dependence with lower redshift. These three redshift samples define our massive quenched galaxy populations in \simbac at these redshifts. Quantitatively, above our $\mstellar\geq10^{10}\mstellar$ mass cut, we find:
\begin{enumerate}
    \item $z=4$ with 9 quiescent and 728 star forming galaxies,
    \item $z=3$ with 86 quiescent and 1975 star forming galaxies,
    \item $z=2$ with 780 quiescent and 4231 star forming galaxies.
\end{enumerate}
Note that these are not the same set of galaxies tracked over time; they are identified independently at each redshift, and indeed we will show that the quenched galaxies don't necessarily overlap.

\begin{table}
    \caption{Alternative main sequence fitting functions.}
    \label{tab:alt_fit}
    \begin{tabular}{cl}
        \hline
        linear & $\log{\rm sSFR}=a\log{\mstellar}+b$\\
        \hline
        quadratic & $\log{\rm sSFR}=a\log{\mstellar}^2+b\log{\mstellar}+c$\\
        \hline
        broken power law & $\log{\rm sSFR} =
        \begin{cases}
            a\log{(\mstellar/M_0)}+c & \mstellar \leq M_0\\
            b\log{(\mstellar/M_0)}+c & \mstellar > M_0
        \end{cases}$\\
        \hline
        smoothed broken & $\log{\rm sSFR} = \log{(A{\rm M}^{\alpha_1}(\frac{1}{2}(1 + {\rm M}^{1/\Delta}))^{(\alpha_1 - \alpha_2)\Delta})}$\\
        power law & where: ${\rm M} = \mstellar/M_0$\\
        \hline
    \end{tabular}
\end{table}

Previous works have also employed linear and broken power law fits to the SFS. We compared alternative fitting functions (equations shown in Table \ref{tab:alt_fit}, adapted from {\sc Astropy.modeling}); the single (linear) power law has two free parameters, the broken power law has three, while the smoothed broken power law has five. The total number of data points after filtering is consistently $\geq 5000$ so the number of degrees of freedom minimally affects the fit. All choices produce reasonable results which become more apparent at lower redshifts, and remain broadly comparable to a flat $sSFR\approx-10$ cut. In detail, we reject a purely linear description of the SFS to retain the SFS turn-down at high masses. We select the quadratic fit because it has the lowest sum of squared residuals from $2\leq z\leq 4$, and due to its slightly greater stability at lower redshifts. However, since the quadratic and both broken power law fits produce results with $\sim90\%$ overlap of identified quiescent galaxies, our results are not sensitive to this choice.

We compare our results to previously reported main-sequences that best match our redshift range, as compiled by \citet{Speagle2014}, a more recent star-forming sequence census by \citet{Leja2022}, and several other determinations. At $z=4$ we find good agreement with previously reported main-sequences. Interestingly at $z=3$ and $z=2$, \simbac matches the newest census, which reports significantly lower specific star formation rates than previous analyses. The main change in the \citet{Leja2022} analysis owes to using flexible star formation histories, which tends to result in a larger underlying older stellar population in massive systems. 

\simbac produces a significant number of quiescent objects at early redshifts, with the $sSFR$ selection method finding $\geq 5$ objects at $z<4.3$, and the earliest such object at $z=5.02$.   This highlights that finding early massive quenched galaxies is certainly possible within modern galaxy formation models, as has also recently been found within the IllustrisTNG~\citep{Donnari2020} and Magneticum~\citep{Kimmig2023} simulations. A number of independent runs produce early quenched galaxies as far as $z=5$, but they fall short of number densities in recent JWST observations \citep{DeLucia2024}. 

Although we don't discuss this further, it is worth noting that compared to \simba, \simbac produces approximately $2\times$ the number of early quenched galaxies at redshifts $z=2-4$. The quiescence is also more pronounced, with many galaxies below the $\text{sSFR}=10^{-11}\msolaryr$ threshold, while most quenched galaxies in \simba lie within the $10^{-10}\msolaryr<\text{sSFR}<10^{-11}\msolaryr$ region; this was particularly evident as \simba failed to match the \citet{Sherman2020} observations which used this stricter definition.  
We speculate that the reasons for this may be twofold:  First, \simbac's updated chemical enrichment model produces less metals overall, resulting in less metal line cooling which is the dominant cooling mode within massive halos; and second, \simbac updates the AGN feedback jet velocity to scale with the black hole mass (to the one-third power) and hence can be more energetic for the very largest black holes.  We leave for the future testing these ideas in detail, and here proceed with focusing only on \simbac.

\section{Quenched fractions in \simbac vs. observations} \label{fraction}

\begin{figure}
    \includegraphics[width=0.95\linewidth]{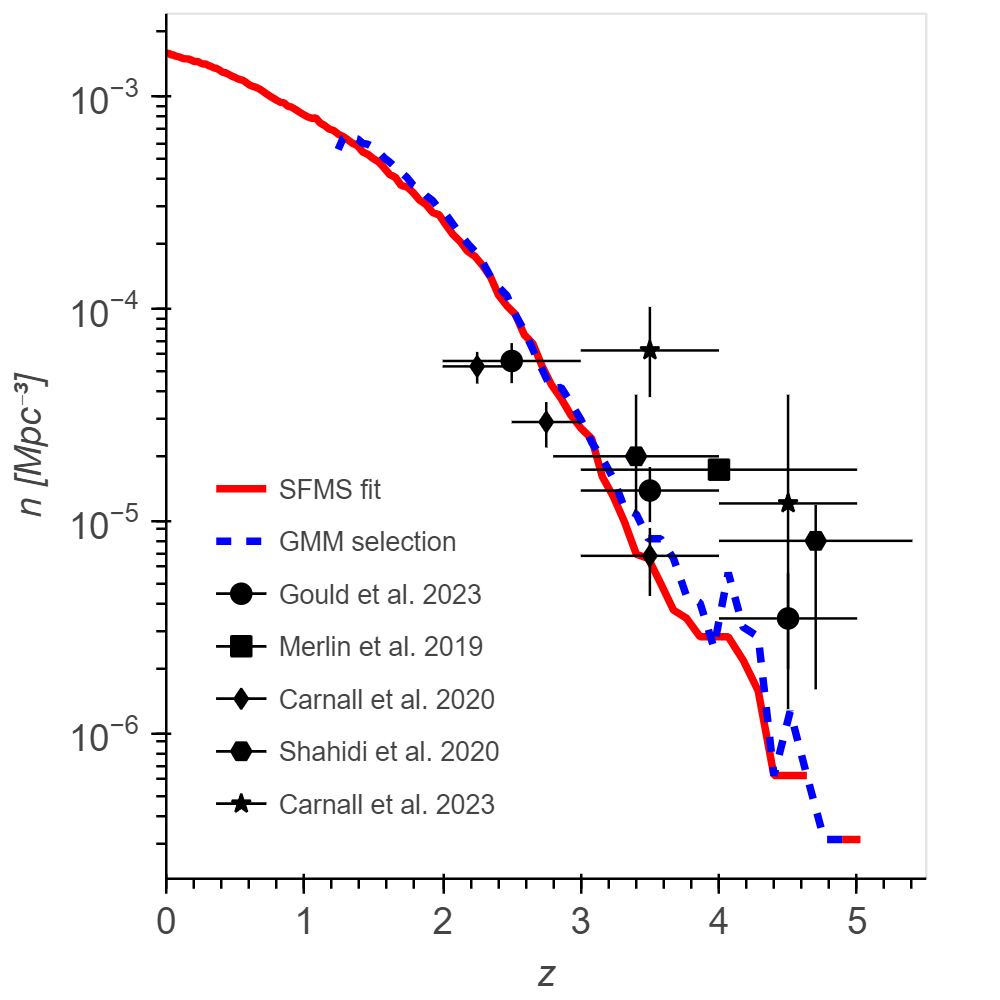}
    \caption{The number density of quenched galaxies as a function of redshift in the range $5.3<z<0$. The red line shows the densities obtained from the SFS fitting process detailed in \S\ref{SFS}. The blue dashed line shows an alternative selection process using NUVU-UV-VJ photometric data with the Gaussian Mixture Model developed by \citet{Gould2023}, with a $M_\star>10^{10}\msolar$ threshold. Black points show observational data for comparison, restricted to $z>2$ for clarity since that is the focus of this work.  The GMM selection (blue line) is an apples-to-apples comparison versus the filled circle data-points from \citet{Gould2023}, which agrees well at $z<4$ but falls short above that.}
    \label{fig:Qdensity}
\end{figure}

\begin{figure}
    \includegraphics[width=0.95\linewidth]{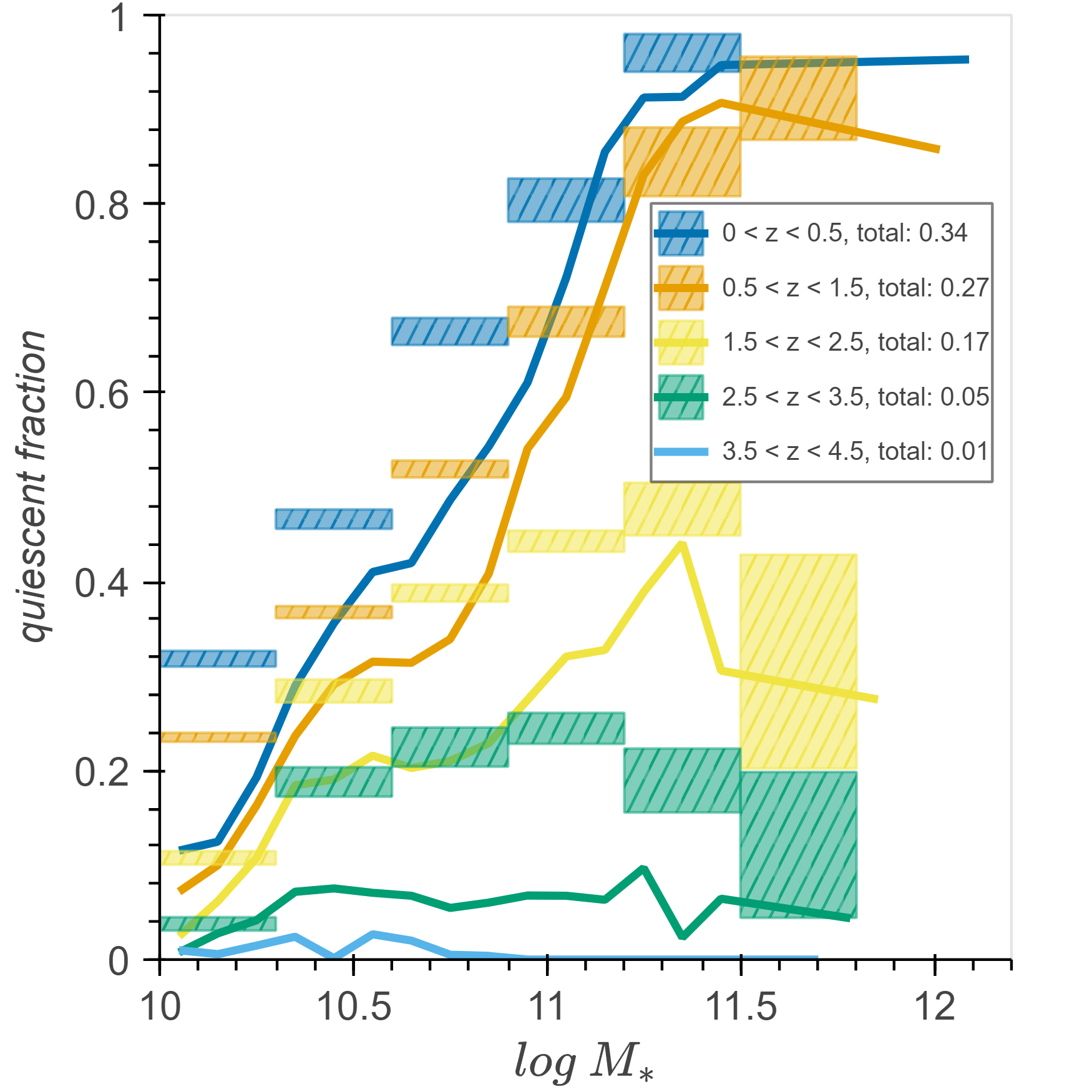}
    \caption{The fraction of quenched galaxies with $\mstellar>10^{10}\msolar$ as a function of stellar mass, within the redshift range $4.5<z<0$. Various colours represent different redshift bins. Solid lines show the predictions from \simbac.  Hatched boxes show data from UltraVISTA DR1 and 3D-HST, adapted from \citet{Martis2016}.
    The observations above $\mstellar=10^{11.5}\msolar$ have been combined into a single bin due to low number of objects.  Note that the observations are selected via a UVJ cut, whereas the simulations use a main sequence-based definition.  Hence this is not an apples-to-apples comparison, but rather highlights various trends in the quenched population from $z\sim 4\to 0$.}
    \label{fig:Qfraction}
\end{figure}

As a baseline test of \simbac and to assess the validity of the modeling, we present the evolution of the quenched fraction of galaxies, first as an overall number density and then broken down by mass.  This fraction is expected to grow over time as more and more galaxies are quenched, owing to slowing accretion and increasing black hole masses that drive quenching via AGN feedback processes. It is also expected to be generally higher in high-mass galaxies, since massive galaxies are more often seen to be quenched.

Figure \ref{fig:Qdensity} shows the number density of quenched galaxies as a function of redshift, comparing \simbac\ to observations as listed (black points). We present the densities obtained with the SFS fit described in \S\ref{SFS} (red line) and an alternative photometric method (blue line) utilising a Gaussian Mixture Model (GMM) developed by \citet{Gould2023} and \citet{Valentino2023}. For more details on the construction of the GMM and it's application to the COSMOS2020 catalog see \citet{Gould2023}.

The two methods agree quite well in the overall number density, mostly within $\sim5\%$ of each other. The identified population also has good overlap of $70-85\%$ for almost all snapshots within the $5>z>2$ range. The GMM selection finds slightly more quiescent galaxies at high redshifts, by including a few that lie just outside our power-law defined SFS. The GMM, fails to detect quenched galaxies at $z<1.5$, which is perhaps not surprising as the GMM was designed to identify quenched galaxies at $z>2$.

Comparing to observations, we first note that the most direct comparison is between the GMM-selected \simbac results (blue line) and the filled circles from \citet{Gould2023} which uses the same selection criterion.  In general, the number densities from \simbac agree quite well with observations at $z\la 4$, which is an important success that has been difficult to achieve in many models~\citep[see][]{Carnall2023}.  However, at $4\la z\la 5$ \simbac falls short by a factor of $\sim\times 2$.  This may be a real discrepancy, but it also may owe to the fact that our finite volume precludes the representation of the very most overdense regions.  As we will show later, the formation of EQGs is tied to living in very overdense regions at earlier epochs, so this may prevent the formation of EQGs sufficiently early on.  In any case, \simbac does a good job reproducing these observations at $z\la 4$, which will be the focus of our analysis in this paper.

We can extend our comparison down to lower redshifts, as well as break it down by stellar mass to provide more detailed tests.  Unfortunately, such compilations are scarce at very high redshifts and subject to uncertainties, but for completeness we present a comparison to the older \citet{Martis2016} compilation of quenched galaxy fractions as a function of $(\mstellar,z)$ from the UltraVISTA and 3D-HST surveys at redshifts $0.2 < z < 3$.

Figure \ref{fig:Qfraction} shows the fraction of quenched galaxies at different redshift intervals from $z\sim 4\to 0$ as a function of stellar mass. The different colours indicate different redshift bins.  The \simbac predictions are shown as solid lines.  In the legend we show the total quenched fraction above $10^{10}\msolar$ in \simbac. Observations from \citet{Martis2016} are depicted as dashed lines with error boxes. Comparing \simbac to data thus involves comparing solid and dashed lines of the same colour.  Here, the observational selection has been done via a UVJ cut, while the simulation results are presented for our main sequence criterion.  Unfortunately, a more direct comparison is not straightforward, because reproducing UVJ diagrams in simulations is fraught with peril owing to the sensitivity to choices such as SPS and dust models; see \citet{Akins2022} for a full examination of these issues in \simba. 

Overall the quenched fraction increases monotonically with time, as expected.  At $z\sim 4$, the overall massive quenched fraction is just 1\%, but by $z=2$ it reaches 17\%, and by $z=0$ it is 34\% of all galaxies with $\mstellar>10^{10}\msolar$.  This shows the accelerating buildup of massive galaxies towards later cosmic epochs, also seen in Figure~\ref{fig:Qdensity}.

Comparing to observations, \simbac tends to fall short by $\sim\times 2$ in the fraction of early quenched systems, although there is good agreement for most massive galaxies at all redshifts.  \citet{Akins2022} broadly found that the UVJ cut picks out galaxies with sSFR approximately below our cut at $z\sim 2$, so this deficit may be real.  However, \citet{Akins2022} also found that there is significant scatter across the UVJ boundary, which may cause UVJ to overestimate the quenched fractions at high redshifts, because there are so many more star-forming systems to scatter into the quenched region than vice-versa. Given the various issues with UVJ selection, at this point we simply note the broad agreement with some shortfall. 

The trends with stellar mass are also quite interesting.  \simbac predicts that at $z\ga 1$, the increase of quenched fractions with $M_\star$ is curiously non-monotonic. This means that, particularly at early times, \simbac's quenched galaxies are not necessarily the very most massive systems in the universe.  This is an interesting prediction that is somewhat counter-intuitive, but arises because (as we will show in \S\ref{Mean evo}) AGN jet feedback turns on around these masses, and the galaxies are still small enough that such energetic feedback can have a major effect on the galaxy, unlike in more massive galaxies that have larger gas reservoirs and more infall.  Additionally, the lack of quenching at high masses is an important reason why \simba yields sufficiently high sub-millimetre galaxy counts to match observations~\citep{Lovell2021}, unlike other hydrodynamic galaxy formation simulations that fall short in part because many of their most massive galaxies at $z>2$ are quenched.

Remarkably, the observations from \citet{Martis2016} also find that at $z\ga 2$ the most massive galaxies are less frequently quenched than slightly less massive systems, with a peak in the quenched fraction at $\sim 10^{11}\msolar$ that increases with time. This is qualitatively similar to what is predicted by \simbac. Hence, there is some observational evidence for the non-monotonic trend with $\mstellar$ seen in \simbac, although in both the simulations and the UltraVISTA DR1 and 3D-HST are subject to small number statistics above $\mstellar>10^{11.5}\msolar$ ($N \la 20$ in \simbac) at high redshift. Larger and more robust samples forthcoming from wide-area surveys will be needed for a more definitive test.

Overall, \simbac produces a sizeable sample of early quenched galaxies, with number densities in agreement with observations, though falling somewhat short at $z\ga 4$, and quiescent fractions that are broadly in agreement with data though may also fall somewhat short. Curiously, \simbac predicts a non-monotonic trend with $M_\star$ in the quenched fractions, with uniform growth at $z\ga 2.5$ and a quickly growing peak at $\sim 10^{11.4}\msolar$ afterwards; currently available observations at $z\ga 2$ suggest a qualitatively similar trend, albeit with poor statistics. We conclude that \simbac represents a viable platform with which to investigate the nature and evolution of EQGs, which is what we do next.

\section{Distinguishing Features of Early Quenched Galaxies}\label{sec:UMAP}

\begin{figure*}
    \includegraphics[width=\linewidth]{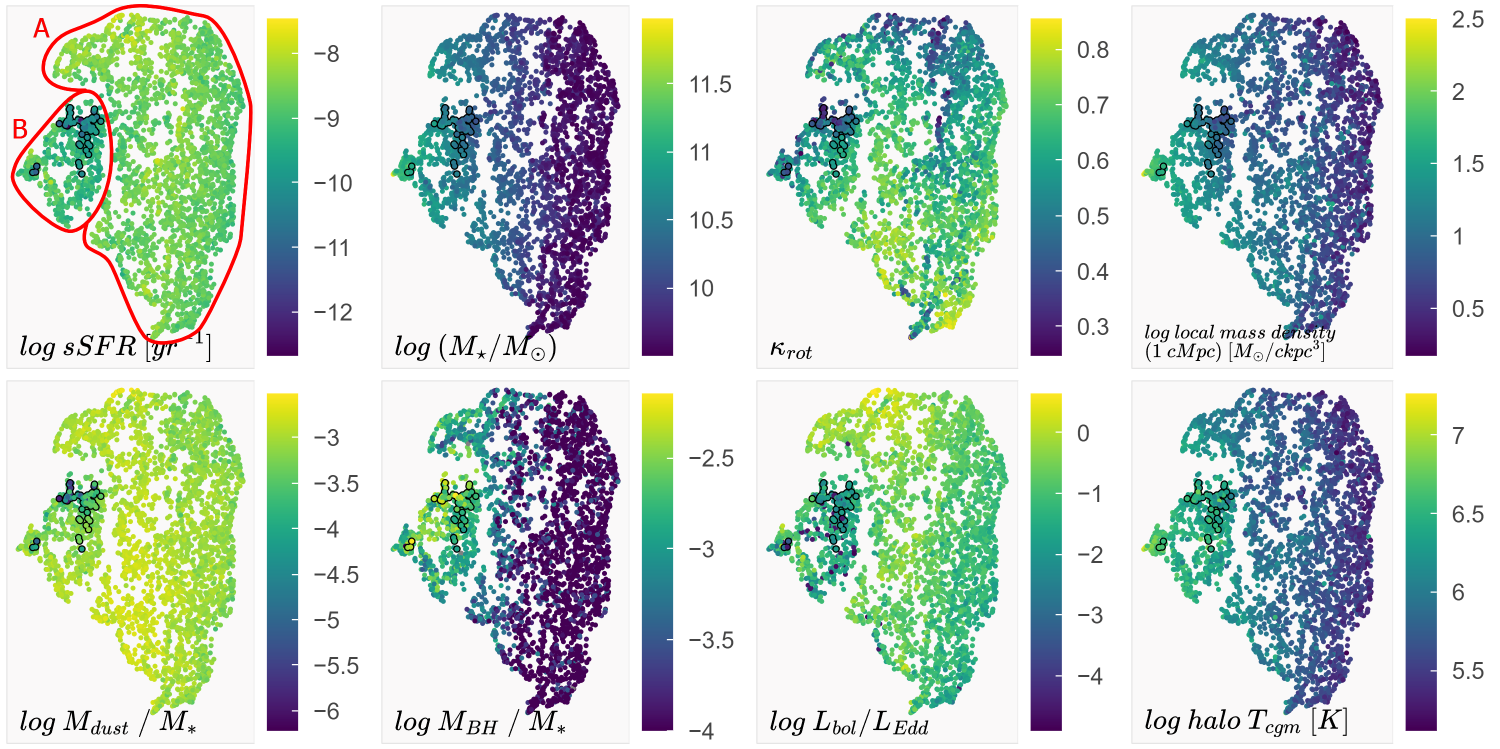}
    \caption{UMAP embedding plot of $z=3$ massive central galaxies ($\mstellar>10^{9.5}\msolar$) coloured by chosen distinguishable features, labelled in bottom left corner of each panel. EQGs are highlighted with a black outline. The embedding forms two distinct clusters A and B, as shown in the sSFR panel. Cluster A seems to be comprised of primarily galaxies firmly on the SFS, while cluster B contains higher mass galaxies with much larger black holes and slightly lower sSFR. EQGs are contained in the top region of cluster B. They have unusually large black holes with $\mbh/\mstellar\gtrsim10^{-3}$ are primarily elliptical with $\kappa_{\rm rot}\lesssim0.4$ and are easily identified by very low sSFR and dust mass. No further separation of cluster B indicates no categorical differences between EQGs and the rest of the cluster, even when considering the local neighbourhood.}
    \label{fig:UMAP}
\end{figure*}

\begin{figure*}
    \includegraphics[width=\linewidth]{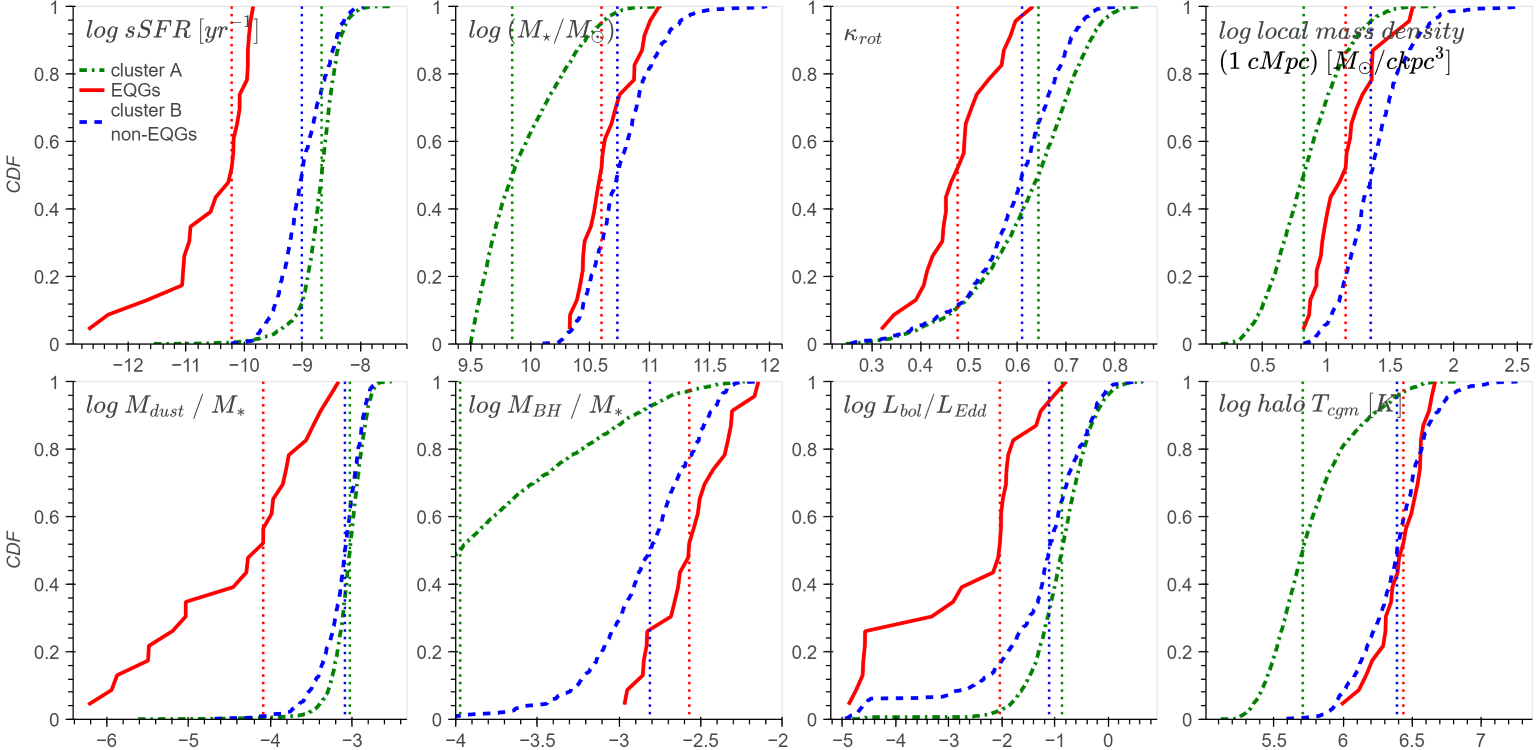}
    \caption{Distribution of EQGs (red solid line) against other cluster B galaxies (blue dashed line) and cluster A galaxies (green dot-dashed line) for chosen distinguishable features. Cluster B galaxies are clearly distinct from cluster A, especially in $\mstellar$ and $f_{\rm BH}$. According to two sample K-S tests, EQGs form a separate population from other cluster B galaxies with at least $\alpha=0.05$ confidence for all properties except halo $T_{\rm cgm}$. The vertical dotted lines denote the median of the respective distributions.}
    \label{fig:Distribution}
\end{figure*}

In this section, we ask some basic questions about the quenched and non-quenched early galaxy population:  What are the physical properties of EQGs? What distinguishes them from non-quenched massive galaxies at that epoch? For illustration we consider the \simbac EQGs selected at $z=3$, as they are qualitatively similar for $z\sim 2-4$.  We further examine evolutionary trends in \S\ref{Progens}.

\subsection{UMAP analysis} 

To elucidate trends and gain intuition, we begin by employing a general purpose visual classification approach using UMAP \citep[Uniform Manifold Approximation and Projection,][]{UMAP2020}, -- a dimension reduction algorithm based on manifold learning techniques and topological data analysis -- to highlight meaningful trends within the galaxy population. Other general purpose approaches exist, such as principal component analysis (PCA; for a review see \citealt{Jolliffe2016}) to find new uncorrelated variables based on linear combinations of initial variables or t-SNE \citep[t-distributed Stochastic Neighbor Embedding,][]{Vandermaaten2008}, an unsupervised non-linear dimensionality reduction algorithm. For this analysis, we choose UMAP due to its effectiveness at extracting and separating small populations distinct from a larger dataset \citep{Packer2019, Cao2019}. For a thorough comparison of UMAP to different dimension reduction algorithms, see \citet{UMAP2020}.

UMAP reduces dimensionality and preserves information by training a machine learning algorithm on the topological structure of the data, and finding a low-dimensional embedding that retains the topological structure of the initial dataset. The final result is a projection into a lower-dimensional space that preserves distance based on a chosen metric. 

The position of data-points in the UMAP embedding is arbitrary, and roughly corresponds to proximity within the high-dimensional space spanned by all the input parameters. The precise values within the 2-dimensional embedding represent some complex multi-dimensional categorical distance which is not meaningful in and of itself.  Hence, our discussion will be focused primarily on identifying regions of interest, where distinctive groups of points can be interpreted by correlating them with specific galaxy properties. These groups can form separate clusters implying categorical differences with other groups, or by forming tight substructure within a larger cluster, indicating existence of strong similarities. In this analysis we apply UMAP to the \simbac $z=3$ galaxy catalogue to determine significant features of early quenched galaxies which distinguish them from other massive galaxies.

We use 62 galaxy physical properties available in the \caesar catalogue as input for UMAP; a full list is available in Appendix A. In this analysis we focus on a physically relevant subset of these properties, the importance of which we establish visually from the UMAP embedding and built-in diagnostics. We limit the input dataset to massive central galaxies i.e. galaxies with highest $\mstellar$ within their halo and with $\mstellar>10^{9.5}\msolar$. Using the same stellar mass cut as for selecting EQGs ($\mstellar>10^{10}\msolar$) yields a similar result, but a much smaller dataset (1785 galaxies vs 3748 with the lower cut). We rescale the input data for comparability to the (1,99) quantile range to limit the influence of outliers and to account for most features not being normally distributed.

For the selection of the quantile range and UMAP hyperparameters we utilise AlignedUMAP, which provides aligned embeddings of a series of datasets by constraining the data-point locations across multiple embeddings during the optimization process. We use $\tt{alignment\_window\_size=1}$ and $\tt{alignment\_regularisation=10^{-3}}$ to minimally influence the embeddings through alignment. Varying the $\tt{n\_neighbors}$, $\tt{min\_dist}$ and $\tt{repulsion\_strength}$ hyperparameters and the quantile range used for data rescaling results in qualitatively equivalent embeddings for $10<\tt{n\_neighbors}<50$, and quantile ranges broader than $(15,85)$, with $\tt{min\_dist}$ and $\tt{repulsion\_strength}$ having minimal influence on the final embedding.  All UMAP models are trained with $\tt{random\_seed}=1$. In the end we choose $\tt{n\_neighbors}=15$, $\tt{min\_dist}=0.1$ and $\tt{repulsion\_strength}=2$, which were chosen by inspection to yield the most distinctive cluster within which the quenched massive galaxies are congregated. To demonstrate this, we present UMAP embeddings for the range of explored alternative hyperparameter selections in Appendix A.

Figure \ref{fig:UMAP} shows the UMAP embedding plot.  Each panel shows the same underlying UMAP, coloured based on specific features (named at the lower left of each panel), with colour bars to the right of each panel showing the values for their respective quantities. We present 8 panels to show expected important properties of early quenched galaxies: sSFR, stellar mass ($\mstellar$), morphology using the fraction of kinetic energy in rotation ($\kappa_{\rm rot}$) defined by \citet{Sales2010}, local environment using the local mass density within $1\;\mpccm$, dust mass using $f_{\rm dust}=\frac{{\rm M}_{\rm dust}}{\mstellar}$, central black hole mass using $f_{\rm BH}=\frac{\mbh}{\mstellar}$, Eddington ratio ($f_{\rm Edd}=\frac{L_{\rm bol}}{L_{\rm Edd}}$) and halo circumgalactic medium (CGM) temperature $T_{\rm cgm}$. We compare to a newer morphology estimator ($\kappa_{\rm co}$) defined by \citet{Correa2017} which restricts the calculation to only star particles corotating with the galaxy and find similar relative differences between $\kappa_{\rm rot}$ and $\kappa_{\rm co}$.

We identify two clusters labelled A and B in the sSFR panel. The EQG point are highlighted with black outlines, and they all lie within cluster B concentrated toward the top of the cluster. Hence, UMAP seems successful at classifying EQGs -- they form a coherent group, somewhat distinct from most other galaxies, but sharing key categorical similarities with some of them.  We will argue that a large fraction of cluster B galaxies are ones which have temporarily quenched before $z=3$ or will quench before $z=2$, but otherwise share many characteristics with $z=3$ EQGs.

Figure \ref{fig:Distribution} shows the cumulative distribution functions of EQGs (red solid line), cluster B non-EQGs (blue dashed line) and cluster A galaxies (green dash-dotted line) for the same properties as the UMAP embedding. We use two sample K-S tests to bolster the visual intuition of splitting $z=3$ massive galaxies into these three categories. We see the underlying distributions are well reflected in the UMAP embedding – cluster B galaxies and especially EQGs are strongly distinct from cluster A galaxies, while the two groups within cluster B are much more similar. 

In the following subsections we consider each physical property presented in each panel in order to obtain a qualitative sense of what properties are common among EQGs in \simbac, and how they differ from properties of the bulk of the galaxy population, split into cluster A and cluster B non-EQGs. We further explore the evolution of these three groups in \S\ref{Progens}.

\subsection{Specific star formation rate}
We first consider sSFR (upper left panel of Figure  \ref{fig:UMAP}), which is of course a defining feature of EQGs.  Hence partly by construction, sSFR provides a clear distinction of the substructure of cluster B, which is divided into the region populated by EQGs and surrounded by similar but non-quenched galaxies.

The primary difference between the clusters is lower sSFR in cluster B: $sSFR\lesssim9.5\msolaryr$, while in cluster A: $sSFR\gtrsim9\msolaryr$.  The EQGs all lie in a distinct region at the top of cluster B, indicating some common features across all early quenched galaxies with significant similarity to the rest of cluster B.  Meanwhile, cluster A contains the large majority ($82\%$) of galaxies all of which lie on the galaxy main sequence.

Other than the significantly lower sSFR among EQGs, both clusters form smooth gradients in sSFR, with no clear discontinuity between the two clusters.  Hence from this plot we suggest that the EQGs are the onse subset of galaxies that occupy a distinct region in sSFR space, with the rest forming a continuum with broadly similar sSFRs.

\subsection{Stellar mass}

The second leftmost panel in the upper row of Figure \ref{fig:UMAP} shows the galaxies colour-coded by $\mstellar$.  Cluster B contains the highest $\mstellar$ galaxies from $z=3$ with on average lower sSFR, coinciding with the very massive galaxy population slightly below the SFS visible in Figure \ref{fig:SFSfit}. The transition between clusters A and B is smooth in stellar mass, so while galaxies in cluster B are generally more massive, it is not a distinguishing factor between the two clusters. 

The EQGs lie in cluster B, but they are not obviously separated in $\mstellar$ from the non-quenched galaxies in cluster B.  In particular, the region of lowest sSFR (top of cluster B), where EQGs lie and the region of highest stellar mass (bottom left of the cluster) do not strongly overlap.  This indicates that at $z=3$ EQGs are not well described as the most massive galaxies at that epoch.  This is consistent with the impression from Figure \ref{fig:SFSfit} that the EQGs do not preferentially occupy the most massive end of the diagram.

\subsection{Morphology}

While \simbac has relatively low resolution so we cannot accurately measure detailed morphological characteristics, we have found that the fraction of kinetic energy in rotation \citep[$\kappa_{\rm rot}$;][]{Sales2010} represents a good proxy for whether a galaxy is visually disk-dominated or spheroidal-dominated~\citep{Kraljic2020}.  $\kappa_{\rm rot}\sim1$ corresponds to discs with perfectly ordered rotation and $\kappa_{\rm rot}\ll1$ for non-rotating systems. We have found that galaxies with $\kappa_{\rm rot}>0.7$ and $\kappa_{\rm rot}<0.5$ can be interpreted as spiral and elliptical, respectively.

Cluster B houses both elliptical and spiral galaxies according to this definition. There is a strong gradient in $\kappa_{\rm rot}$ across cluster B, with the ellipticals on the left and top, and the spirals towards the bottom.  This also broadly correlates with the trend in $\mstellar$, where the ellipticals tend to be more massive.  Hence, \simbac reproduces the long-observed trend that more massive galaxies tend to be more spheroidal, even at a broadly similar sSFR.

The EQGs themselves tend to be ellipticals.  Hence, although they are not necessarily the most massive, they do tend to be more spheroidal.  This suggests that mass is not the only determinant of the level of rotational support, but quenching also plays a role in morphological transformation or the other way round, morphology changes lead to quenching.

\subsection{Local environment}

At low redshifts, there is a strong colour-density relation for central galaxies, with quenched galaxies preferentially residing in dense environments~\citep{PengLilly2010}.  At $z\ga 1$ this is less clear, but the densest protoclusters still seem to host a nascent red sequence of quenched galaxies~\citep{Overzier2016}. Hence, it is interesting to ask whether \simbac predicts that EQGs live in denser environments.

We measure the local environment via the mass density within a tophat spherical aperture to obtain the local mass density\footnote{Using galaxy counts instead of mass gives similar results but the results are noisier.}.  

In Figure~\ref{fig:UMAP}, upper right, we show the $z=3$ UMAP colour-coded by environment, choosing an aperture of 1~\mpccm, although as we show next the results are not sensitive to this choice.

In general, galaxies in group B tend to be in higher density environments than group A, as expected from the fact that it contains more massive galaxies.  Hence \simbac unsurprisingly predicts a strong mass-density relation.
Even within group B, there is a substantial trend that high-density galaxies live towards the left of the group, which corresponds to their higher mass.  

The EQGs, however, live at the upper end of group B, so they do not necessarily live in the densest environments.  Instead, they appear to live in intermediate environments, somewhat denser than the overall population but not clearly in the protoclusters.  Such environments might be ideal for merging which could drive quenching~\citep{Hopkins2008}, being dense enough for there to be close galaxies, but the halos are not so massive as to have overly high relative velocities.  However, we investigated further and found that there was no correlation of EQGs with recent merger activity.  It appears that environment, just like stellar mass, is not the primary determinant of why these particular galaxies quench at this epoch.

\subsection{Black hole fraction}

In modern galaxy formation models, quenching is often driven by the cumulative action of AGN feedback~\citep[e.g.][]{SomervilleDave2015}, which adds energy to the surrounding gas in order to shut off the cooling flow from circum-galactic gas.  In this case, one might expect that the cumulative black hole accretion which yields increased overall feedback energy would result in a more massive black hole.  Here we check this in \simbac.  To mitigate the ``larger is larger" effect and elucidate more subtle trends at a given $\mstellar$, we scale the size of the black hole to its host galaxy stellar mass and consider the black hole fraction: $\fbh=\frac{\mbh}{M_{*}}$. 

In cluster A a clear pattern emerges with $\fbh$ increasing with the stellar mass.  This emerges from \simbac's accretion model in which the accretion is increasingly suppressed at low mass in order to mimic the (unresolvable) effect of star formation feedback in the immediate vicinity of the black hole~\citep{Angles2017b}. Also, cluster A clearly has lower $\fbh$ than cluster B, with $\fbh\ga 10^{-3.0}$ in cluster B and $\fbh\la 10^{-3.5}$ in cluster A. 

For cluster B, however, there is a less pronounced trend of $\fbh$ with $\mstellar$.  However, in detail a trend still emerges: The upper leftmost region has $\fbh\ga 10^{-2.5}$, while the rest of the cluster has smaller $\fbh$.  Indeed, this is exactly the same region of the diagram where the EQGs lie. This suggests that, as opposed to stellar mass or environment, anomalously early and rapid black hole growth is a significant factor for early galaxy quenching.  We will explore this further in \S\ref{Typical}.

\subsection{Dust mass}

\simbac includes a model to form, grow, and destroy dust on-the-fly, as in \simba~\citep{Li2019}.  As galaxies quench, this causes the dust to be destroyed, primarily via sputtering in hot  gas~\citep{Donevski2020,Whitaker2021}, which lowers the dust-to-stellar mass ratio $\fdust$.  An interesting question is whether this process still operates in EQGs, since in the early universe the gas is denser and it is more difficult to sustain hot gas.

The leftmost panel of the lower row shows that indeed this effect is still very strong at high redshifts.  Overall, there are mild variations in $\fdust$ for most galaxies, but the EQGs have anomalously low $\fdust$ compared to all other galaxies, even those in cluster B.  While this is likely to be an effect not a cause of early quenching, it suggests that the cool gas must be heated and/or removed in order to yield an EQG.

\subsection{Eddington ratio}

We further consider the impact of early black hole growth on EQGs by considering black hole accretion using the Eddington ratio $f_{\rm Edd}=\frac{L_{\rm bol}}{L_{\rm Edd}}$, where $L_{bol}$ is the bolometric luminosity of the AGN and $L_{\rm Edd}$ is the Eddington luminosity.

Cluster B has a very large spread $10^{-5}\la f_{\rm Edd}\la 1$ with no clear distribution within the cluster, but a slight tendency for higher $f_{\rm Edd}$ in the most massive galaxies. EQGs follow a very different trend in their distribution that we will discuss in the following section, where they follow one of two regimes: very low ratios $f_{\rm Edd}\la10^{-4}$, which implies little accretion in those galaxies and indicate that the jet-mode AGN feedback is the key for the quenching of EQGs, and higher ratios $f_{\rm Edd}\ga10^{-2}$, which are still an order of magnitude lower than the median of other cluster B galaxies, but indicate that some galaxies can recover the material lost during quenching, as we will discuss in \S\ref{Mean evo}.

We further examined whether the black hole accretion rates are different in the EQG and non-EQG populations.  The differences are less prominent than for $f_{\rm Edd}$; while the Eddington ratios are lower, the black hole masses are higher, resulting in less differences in the accretion rate.  However, the accretion rates are indeed lower at $z=3$ (less so at $z=2$) for EQGs, which is somewhat counter to the results from IllustrisTNG and EAGLE in \citet{Bluck2023}, who find that instantaneous black hole accretion rate has little predictive power for quenching.  \citet{Bluck2024} argues instead that central stellar potential provides a better indicator of quenching, as a proxy for black hole mass.  Our results with \simba\ agree that high black hole mass is more strongly correlated with quenching, but we do expect that higher-$z$ EQGs will have somewhat lower black hole accretion rates along with lower Eddington ratios.

\subsection{Halo CGM Temperature}

Recent AGN feedback is closely linked to the CGM temperature of the halo, as strong energy outflows heat up the gas surrounding a galaxy. Given the expected importance of AGN during quenching, halos containing EQGs can be predicted to have higher CGM temperature.

While galaxies in cluster B do have higher CGM temperature with $T_{\rm cgm}\ga10^{6}K$, EQGs are not associated with increased temperature in comparison to other cluster B galaxies. This suggests that AGN feedback is not stronger in EQGs than other early massive galaxies, but is sufficient to quench smaller massive galaxies, which we will explore in \S\ref{Mean evo}.

\subsection{Comparison of EQGs to the non-EQG population}
We now consider the cumulative distribution function of the above properties, in order to more quantitatively explore the differences between EQGs, Cluster B non-EQGs, and Cluster A galaxies.

Figure \ref{fig:Distribution} shows histograms of the 8 physical quantities shown in Figure~\ref{fig:UMAP}, for EQGs (red, solid line), non-EQGs from cluster B (blue, dashed line) and cluster A (green, dot-dashed line) for comparison. We will discuss primarily cluster B galaxies from the UMAP projection, to focus on broadly similar galaxies, as cluster B contains all EQGs. The median value of each relation is shown as the vertical dotted line.

The sSFR histogram shows, as expected, that the EQGs have significantly lower sSFR.  The median value of the EQGs is a bit below $10^{-10}$yr$^{-1}$, showing that the majority of quenched galaxies are not completely devoid of ongoing star formation.

The $\mstellar$ histogram shows little difference between the EQG and non-EQG populations, with only 0.1~dex difference in the median values. This is also the case for the halo mass and stellar-to-halo mass ratio (not shown).  This confirms our visual impressions from the UMAPs that stellar or halo mass are not the primary determinant for whether a galaxy quenches.  We note that because the stellar and halo mass distributions are similar, we could have instead chosen to select above some $\mstellar$ or ${\rm M}_{\rm halo}$ threshold and the results would have been similar.

The environments of EQGs and non-EQGs are significantly different, but in the opposite sense of what might be naively expected:  EQGs lie in lower density environments than non-EQGs.  This is in contrast to what is observed (and, as we will show below, predicted) at low redshifts where quenched galaxies live in higher density environments.  We will explore further the physical interpretation of these trends with environment in \S\ref{fraction}.

The black hole mass fraction $\fbh$ shows a modest but significant difference of 0.4~dex in the median values. Together with the complete lack of EQGs with $\fbh<10^{-3}$ provides a clear case for the importance of rapid early black hole growth and subsequent AGN activity for early galaxy quenching.

$\fdust$ shows dramatic differences between EQGs and non-EQGs. We do not show it, but also the $H_2$ and \ion{H}{i} mass fractions also show strong differences, in the sense that EQGs have much lower cold gas contents; this follows expectations for galaxies that are selected to be with lower star formation.  These properties are likely not to be causal but emergent from the various physical processes that quench the galaxies.  The lower dust contents are likely a reflection of the heating or removal of cool gas, which causes spallation of dust.

The morphological differences seen in $\kappa$ are somewhat trickier to understand.  This could arise from minor mergers, which can disrupt disks, which cannot regrow or re-settle owing to AGN feedback.  It could in fact be predictive in the sense that EQGs have undergone more mergers.  However, a more likely scenario is that it is emergent from quenching, which prevents the accretion of colder gas settling into a rotating disk configuration.  As a result the high merger rates at high redshift are able to disrupt the rotation without regrowing a disk, resulting in low rotational support.

The histograms of Eddington fractions are also substantively different, with the EQGs showing an order of magnitude lower median $\fedd$ than the rest of the Cluster B population.  Importantly, the median value among EQGs lies below the threshold of $\fedd=0.02$ which in \simbac signals the onset of full-speed AGN jet feedback along with AGN X-ray feedback.  This strongly suggests that EQGs distinguish themselves from otherwise similar non-quenched galaxies by having more energetic AGN feedback.  In \S\ref{Mean evo} we will show this is indeed the case.

Finally, we examine whether EQGs live in hotter environments.  We measure this using $T_{\rm cgm}$, which is defined to be the mean mass-weighted temperature of all gas within the EQG's halo that is below \simbac's star formation threshold density of $n_H<0.13$~cm$^{-3}$.  One might expect that halo heating from AGN feedback causes starvation of the central galaxy that leads to quenching~\citep[e.g.][]{Croton2006}, but the histograms show no difference between EQGs and non-EQGs within Cluster B.  We do not show halo mass among these properties, but halo mass is also negligibly different, suggesting that the CGM temperature is set primarily by halo mass rather than any effects of feedback.  Hence if AGN feedback is the cause of quenching as indicated from the $\fedd$ histograms above, then it is not doing so by heating the entire halo.

Overall, our UMAP analysis suggests that compared to the overall population of galaxies, EQGs tend to be more massive, live in denser environments, and have lower rotational support.  However, if we focus on just the galaxies in Cluster B which all broadly share these overall characteristics, then EQGs stand out for having high $\fbh$ and low $\fedd$, but not for having higher stellar masses or living in hotter or denser environments. They also have somewhat lower rotational support and lower dust masses, which are likely to be caused by the quenching process but represent interesting testable predictions for future observations.

\section{Progenitors and Descendants of EQGs} \label{Progens}

\begin{figure*}
    \includegraphics[width=\linewidth]{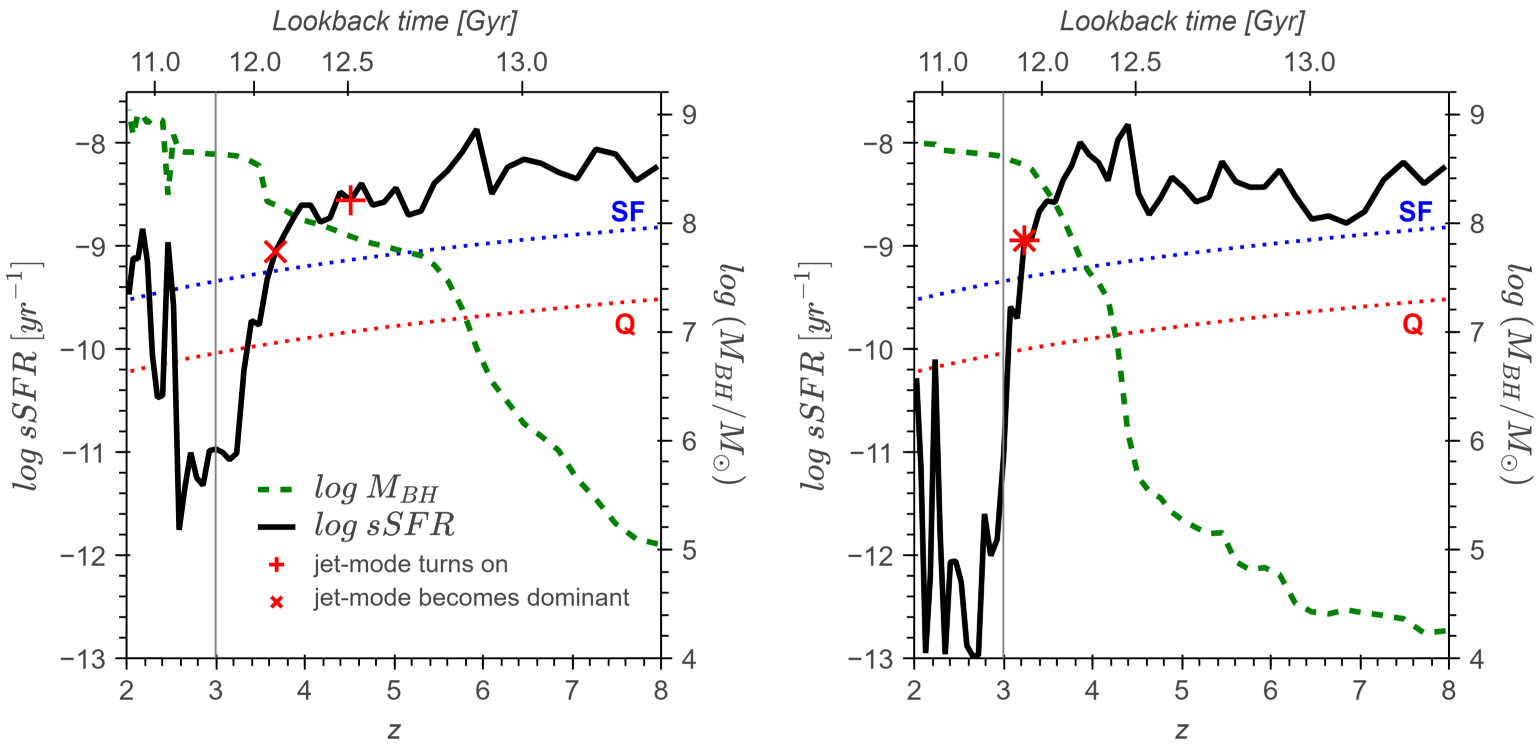}
    \caption{Evolution of selected physical properties of two among the most massive EQGs detected at redshift $z=3$. The black solid line shows the sSFR, green dashed line shows the mass of the central black hole and the blue and red dotted lines show the threshold for rejuvenation (labelled SF) and quiescence (labelled Q), defined as $\text{sSFR}<1\tau_H$ and $\text{sSFR}<0.2\tau_H$ respectively, following \citet{Rodriguez2022}. We observe a common evolutionary trajectory: EQGs undergo early black hole growth with the central black hole transitioning to jet mode near redshift $z=4$ as the growth of the black hole slows down. We denote the transition to jet mode with a red plus and the red cross shows the moment when jet mode feedback accounts for $>50\%$ of the total black hole feedback energy. Due to the strong AGN jet activity, the galaxy rapidly quenches to usually below $\text{sSFR}=10^{-11}{\rm yr}^{-1}$ often chosen as a quiescence threshold for observations. Some galaxies experience rejuvenation (left) after a period of quiescence lasting $200-1000$ Myr.}
    \label{fig:typical}
\end{figure*}

\begin{figure*}
    \includegraphics[width=0.95\linewidth]{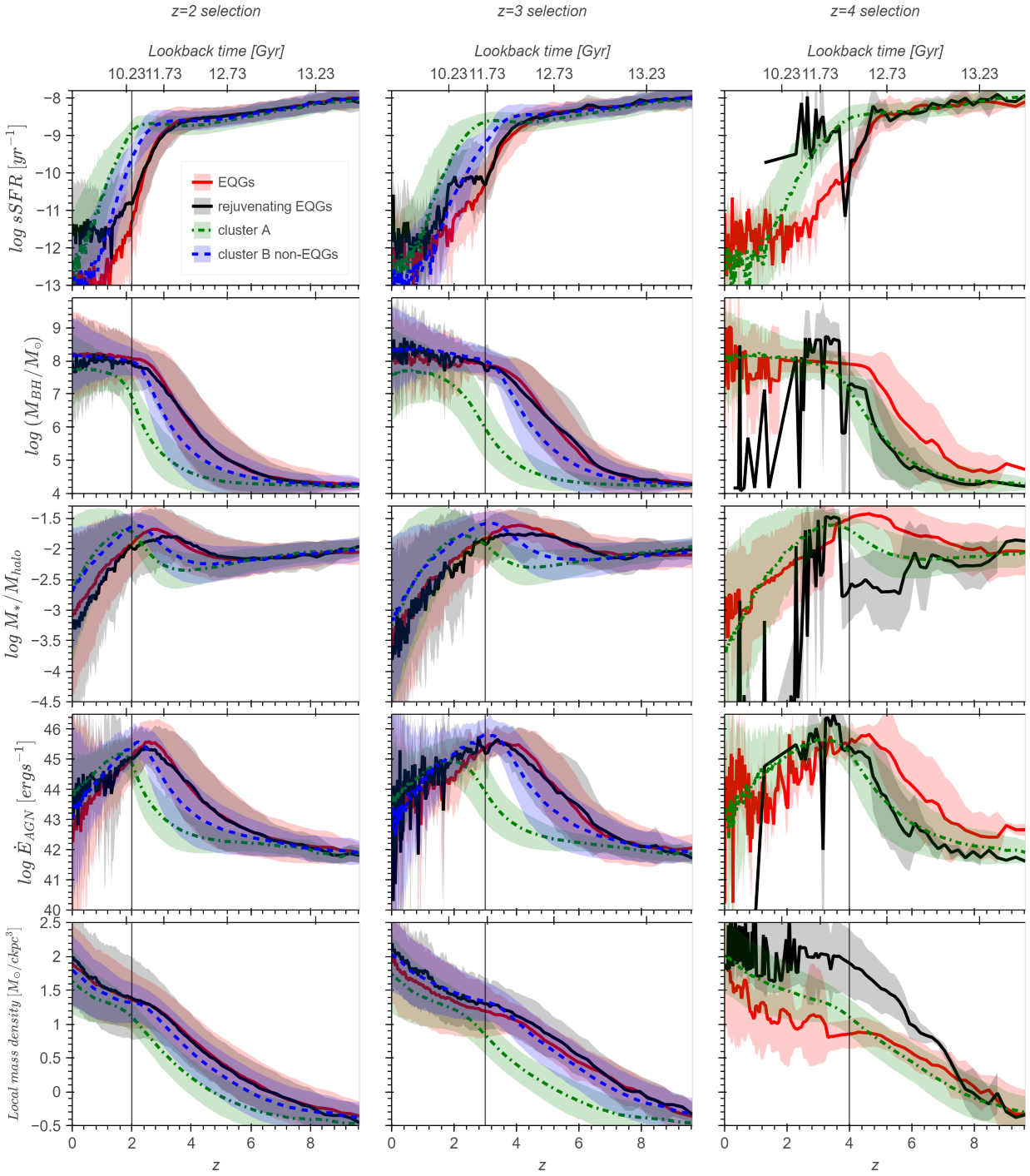}
    \caption{Redshift evolution between $z=10$ and $z=0$ of chosen physical properties for the $z=2$, $z=3$ and $z=4$ selections of massive quiescent and star forming galaxies in \simbac. The quiescent population is split into rejuvenating (black, solid) and non-rejuvenating (red, solid) based on crossing the star formation threshold of $\frac{1}{\tau_{H}}$ for at least two snapshots after initially quenching. Star forming galaxies are divided into cluster A (green, dot-dashed) and cluster B (blue, dashed) based on the UMAP embedding, where cluster B contains EQGs. The $z=4$ embedding does not split into two clusters, therefore all star-forming galaxies are shown as cluster A. We show the mean evolution of the entire population (solid line) with the associated standard deviation (shaded regions). $z=4$ data suffers from low statistics, with only 9 EQGs. Local mass density (fourth row) is follows the previous scale of $1\mpccm$. Overall the sSFR of quiescent galaxies falls below other massive galaxies $0.5-1 Gyr$ before being selected as quenched. They undergo an accelerated growth of the central black hole (second row), likely due to their high density environment. We observe some separation between the rejuvenating and non-rejuvenating populations near $z=6$, respectively remaining in high density environments and falling into low density environments (third row), although the separation is not visible for the $z=2$ selection. Just before quenching, we observe strong AGN jet mode emissions (fourth row), with the $z=2$ population experiencing much stronger AGN activity. Cluster B star-forming galaxies show significant similarities to the EQG trajectories, but slightly delayed.}
    \label{fig:evo}
\end{figure*}

\begin{figure*}
    \centering
    \includegraphics[width=0.95\linewidth]{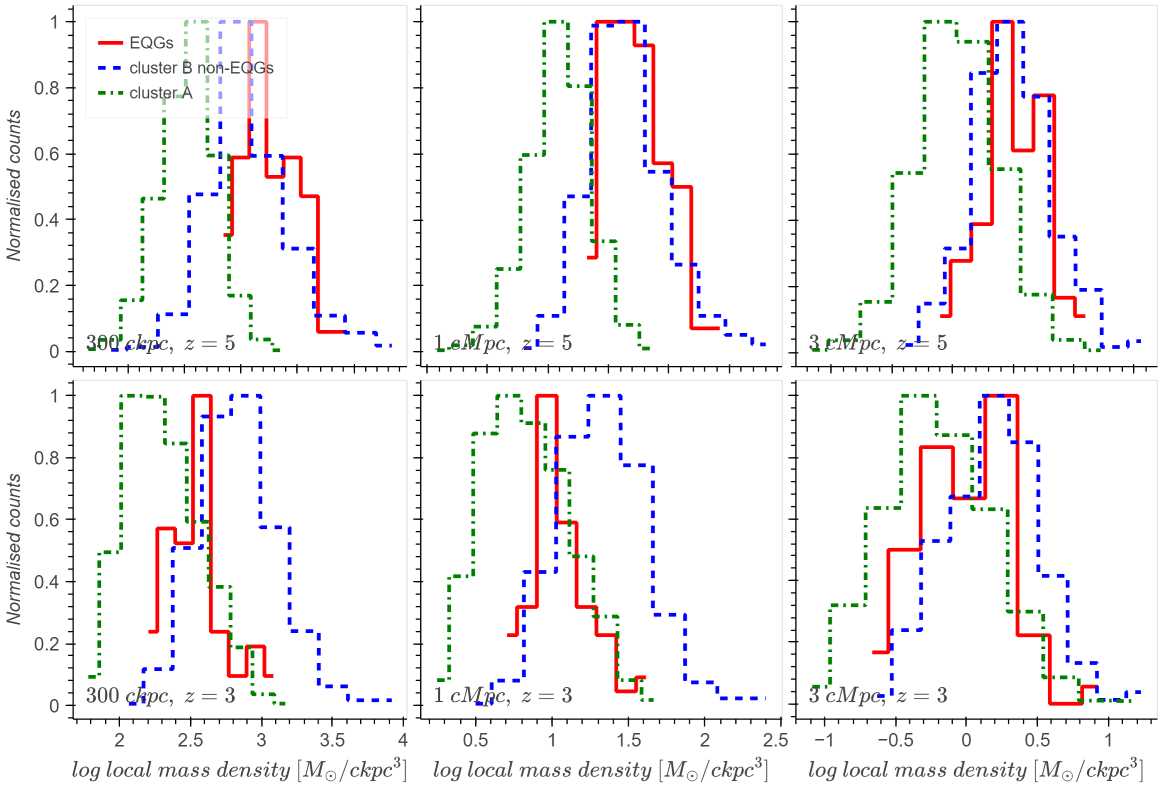}
    \caption{Distribution of $z=3$ (bottom row) and $z=5$ progenitors (top row) of EQGs (red solid line) against cluster B non-EQGs (blue dashed line) and cluster A (green dot-dashed line) galaxies for local mass density at 300 ckpc, 1 cMpc and 3 cMpc. Cluster B galaxies tend to be in higher density environments than cluster A, corresponding to their higher mass. EQGs follow a surprising trend: their progenitors at $z=5$ are in higher density environments especially at lower scales, while at time of identification at $z=3$ they lie in lower density environemnts than other massive galaxies. This could point to the importance of high mass mergers causing early quenching.}
    \label{fig:localmassdens}
\end{figure*}

Having examined the prominent features of quiescent galaxies at the time of selection at high redshift, we now turn our attention to understanding the history and fate of individual EQGs in order to gain insights into the processes responsible for quenching, as well as to understand their evolution to $z=0$.  To do this, we determine the progenitors and descendants of each EQG selected at a particular redshift, by matching up the galaxies in earlier and later snapshots that have sufficient number of star particles in common, then taking a stellar mass-weighted average of those galaxies. In order to have a stable definition of the evolution of descendants of EQGs, we follow \citet{Rodriguez2022} who define star-forming galaxies as having ${\rm sSFR}>\frac{1}{\tau_{\rm H}}$ and quiescent galaxies as those with ${\rm sSFR}<\frac{0.2}{\tau_{\rm H}}$, where $\tau_{\rm H}$ is the Hubble time at a given redshift. With this definition, we can also define a ``rejuvenating" galaxy following a similar scheme as \citet{Rodriguez2022} as one that after initially quenching based on the SFS definition we describe in \S\ref{SFS}, later rises above the star-forming threshold for a number of consecutive snapshots. We discuss the sensitivity to the chosen number in \S\ref{sec:rejuvenation}. 

In the following subsections, we begin with the analysis of the evolution of two typical EQGs detected at $z=3$. The common pattern across all EQGs in \simbac is the very early growth of $\mstellar$ and $\mbh$ with strong AGN jet activity, which peaks during the quenching period. Informed by this, we consider the mean evolution of the entire quiescent population at redshifts $z=2$, $z=3$ and $z=4$ and compare them to the star-forming massive galaxies at those epochs as well as distinguishing rejuvenating vs. non-rejuvenating galaxies. We then quantify the rejuvenating EQG population, and explore sensitivity to our precise definition of rejuvenation.  Finally, to determine if early quenched galaxies retain features which distinguish them from other massive galaxies until today, we examine a UMAP projection of the $z=0$ descendants of EQGs. 

\subsection{Examples of EQG Evolution} \label{Typical}
Figure \ref{fig:typical} presents the evolution from $z=8\to 2$ of two EQGs selected at $z=3$ that illustrate typical trajectories of EQGs in \simbac. The black solid line shows the evolution of sSFR, which can be compared to the dotted curves which show the evolving sSFR thresholds for the star-forming and quiescent populations; when the black curve is above the blue curve, it is star-forming, while it must drop below the red curve to be quenched.  The black hole mass evolution is shown by the green dashed line, with the scale along the right axis.  Finally, we highlight when the black hole passes the minimum mass threshold to allow jet mode of $\mbh>7*10^7-10^8\mstellar$ with a red plus and where the AGN power comprises of at least 50\% arising from AGN jets (which we will refer to as ``jet dominated") with a red cross. A galaxy usually becomes jet dominated right before, and remains as such until quenching. The vertical thin grey line highlights $z=3$ which is the time at which both galaxies are identified as EQGs.

The two examples show representative cases from two groups.  The left panel shows a rejuvenating galaxy, which returns to the main sequence after a brief quiescence period, while the right panel shows a galaxy that is permanently quenched by $z=3$, with minimal star-formation all the way to $z=0$ (though for clarity we only show the evolution to $z=2$).   We will quantify the population fractions within each group in \S\ref{sec:rejuvenation}, but here we focus on these typical cases.

Both groups undergo very similar early evolution.  Broadly, both show early rapid black hole growth reaching $\mbh=10^{8}\msolar$ by $z=4-6$. The growth then slows down with the onset of stronger AGN feedback, with the central BH reaching $\mbh\approx10^{8.5}\msolar$ by the time of quenching, and further growth suppressed by AGN feedback.  The left panel shows an example where at $z\sim 3.5$ the BH undergoes a merger and thus has a rapid increase in mass just before quenching; the right hand example has no such major merger, with rapid early growth driven by accretion. This illustrates that major mergers can be a factor in quenching in some cases, but it is not required.

The sSFR remains stable during the early black hole growth phase at $sSFR\approx10^{-8.5} {\rm yr}^{-1}$. Just prior to $z=3$, the black hole accretion rate slows down, and AGN feedback transitions to jet mode due to the lower Eddington ratio. As the jet mode feedback begins to dominate, the sSFR rapidly declines to $sSFR\approx10^{-11} {\rm yr}^{-1}$, typical for quenched galaxies. The early quenching is fairly rapid, with galaxies completing the transition from star-forming: sSFR$>1/\tau_{\rm H}$, to quenched: sSFR$\sim10^{-11}{\rm yr}^{-1}$ within $\la 100$ Myr (i.e. 2 snapshots). 

It is interesting that in virtually ever case, we find that quenching occurs just after jet mode AGN feedback becomes dominant.  In both cases we present, starting near $z\sim 4-5$ when the hosted black holes reach $\mbh\approx10^{8}\msolar$, the AGN feedback power is $\dot{E}_{jet}\approx 10^{46}$~erg~s$^{-1}$.  However, this does not lead to quenching when such power is ejected in low-velocity radiative mode, only when the high-velocity jet mode turns on.  We further find that in all cases the total energy deposition reaches of order $\sim10^{59}$~ergs prior to quenching, but we could not find a more precise threshold of cumulative AGN energy above which quenching happens.  Hence we attribute quenching to be most closely associated with the onset of jet mode feedback in \simbac, rather than a black hole feedback energy injection rate threshold or a cumulative energy deposition threshold. A recent study on \illustris by \citet{Kurinchi2023} show similar results, with the onset of a kinetic AGN feedback mode and high-redshift environmental factors playing the most significant role for early quenching. We discuss the role of environment in \simbac in section \ref{Mean evo}.

In the context of \simbac's AGN feedback model, this points to two possibilities responsible for early quenching: First, that the jet velocity rather than energy is a critical determinant for quenching, or that the X-ray feedback which is commensurate with jet feedback (and is not enabled when the BH is in radiative feedback mode) is crucial. At lower redshifts, X-ray feedback is critical for removing residual cold gas to produce fully quenched galaxies \citep{Dave2019,Scharre2024}.  The same situation is probably occurring at high redshifts. The spherical nature of \simbac's X-ray feedback which pushes ISM gas outwards is likely important for enabling the kinetic feedback to be effective by lowering the gas density and hence increasing its entropy.  Thus it is probably the combination of jet and X-ray feedback that enacts early quenching.

Right after quenching, there is a brief period of no AGN feedback, indicating no accretion onto the black hole likely due to all available material being removed. The evolution of the two groups diverges after this period of inactivity. In the rejuvenating galaxy case gas begins accreting again and star formation can restart,  enough for the galaxy to cross over the rejuvenation threshold and be identified as a star-forming galaxy. There is a hint of a smaller merger which causes an oscillation and eventual jump in the BH mass, which may represent a merger with a galaxy containing some cold gas.  This reignites jet feedback activity, which eventually quenches the galaxy permanently. Rejuvenation is typically short lived, usually lasting $200-500$ Myr.  Meanwhile, the non-rejuvenating case has a more boring post-quenching evolution, and simply remains quenched with very little BH or stellar growth.

The exact mechanism which differentiates rejuvenating and non-rejuvenating galaxies is unclear from this analysis. Both experience strong AGN jets of the same magnitude $\dot{E}_{AGN}\approx10^{46}$ and associated X-ray feedback, but only some are prevented from restarting star formation. There is some evidence that rejuvenation might be related to mergers post-quenching, but examples that rejuvenate with only accretion also appear. This motivates a statistical approach employed in the following section to consider the mean evolution of each group against each other, as well as against non-quenching early massive galaxies.

\subsection{The Early Growth of EQGs} \label{Mean evo}

In this section we extend on the individual cases in the previous section to present a population-based approach to examining EQG evolution.  We mainly focus on distinguishing between EQGs and other massive non-EQGs (i.e. from cluster A in the UMAP), and we split EQGs into rejuvenating and non-rejuvenating populations as well in order to explore whether there are any global physical quantities which distinguish these.

Figure \ref{fig:evo} shows the mean time evolution of quiescent galaxies selected at $z=2$, $z=3$ and $z=4$ split into rejuvenating (black solid line) and non-rejuvenating (red solid line) sub-populations, compared with other massive galaxies selected at those redshifts, split into cluster A (green dot-dashed line) and B (blue dashed line) based on UMAP embeddings, where cluster B contains all EQGs.  We define rejuvenation as a previously quenched galaxy being star-forming for at least two consecutive snapshots, corresponding to $\ga 50-100$~Myr at these redshifts.  We show the mean specific star formation rate, stellar mass, black hole mass, jet mode AGN feedback energy $\dot{E}_{jet}$ and local mass density within $1\ \mpccm$ to consider potential causes of early quenching and evolution divergent from other massive galaxies. Grey vertical lines mark the time of selection of the given group.

The sSFR of all these groups are comparable until $\sim1$~Gyr before early quenching events, and only then do the EQGs having increasingly lower sSFR over the few hundred Myr prior to the selection redshift.  This is similar for both $z=2$ and $z=3$ selection, while at $z=4$ the small number of EQGs preclude any strong conclusion.  Interestingly, the massive non-EQGs also show a drop in sSFR, but occurring about 1~Gyr behind the EQGs.  Hence EQGs represent the early end of a population that will mostly quench by later epochs.

Looking at the BH mass (second row), this shows differences to much earlier epochs.  The EQGs experience accelerated growth from very early epochs ($z\sim 7$), with black holes growing on average much earlier than in other massive galaxies.  Indeed, the track followed by EQGs shows an offset of a few hundred Myr, similar to the offset in the sSFR track. This highlights the major role of AGN feedback in the early galaxy quenching. 

Again there is little difference between the rejuvenating and non-rejuvenating populations here besides the third column, z=4 selection, in which these rejuvenating galaxies tend to have a later, but much more rapid growth in BH mass before quenching -- jumping from the star-forming curve to the quenched curve. 

The third row shows the stellar-to-halo mass ratio (SHMR).  Here we see another qualitatively different pattern:  Starting at $z\sim 6-7$, the EQGs tend to have high stellar masses for their halo mass, but by the time of selection they are very similar.  After the quenching, this SHMR tends to be higher for these blue lines. This is in agreement with the finding from \cite{Cui2021}, which shows that the earlier quenched red galaxies tend to have lower stellar mass compared to these star-forming ones at the same halo mass.  Again it seems like the non-EQG evolution lags the EQG one by a few hundred Myr.

The fourth row shows the AGN kinetic feedback power being released by the BH.  The evolution is reminiscent of the SHMR, with EQGs having higher AGN feedback power before the selection time and slightly lower after.  The AGN feedback power peaks just before the quenching, driven by the large black holes and dropping accretion rates (as also reflected in the sSFR) within a few hundred Myr of the selection time.  This gives rise to low Eddington ratios that transition the BH into jet mode.  Since \simbac assumes a constant momentum output rate, the power thus scales with the ejection velocity, and hence the high-velocity jet mode tends to dominate the energy budget whenever it is active.  This high energy presumably enacts galaxy quenching.

In summary, comparing the EQG and non-EQG population growth trends, it seems the main discriminating factor is that EQGs experience overall earlier growth of both their BH and stellar mass, even relative to their halo mass. This early BH growth along with the subsequent drop in accretion (and hence SFR) just before quenching resulting in low Eddington ratios that allow the EQG to go into jet mode AGN feedback with its commensurate X-ray feedback, which then quenches the galaxy as illustrated by the examples in the previous section.  The next natural question is, what drives the earlier growth for these specific galaxies?  We examine this question next.

\subsection{The role of environment in EQGs}

The bottom row of Figure \ref{fig:evo} shows the evolution of the mean local environmental density around these galaxy populations.  Here we show the evolution for the 1~cMpc aperture, as before. Note that at $z=2,3$ the red (non-rejuvenating EQGs) and green (rejuvenating EQGs) lines essentially overlap.

At the time of quenching, the difference in environments of EQGs and non-EQGs is minimal, at least at $z=2,3$ where the statistics are good; this confirms what we saw in \S\ref{sec:UMAP}.  However, at any earlier epoch back to $z\sim 8$, EQGs clearly live in more dense environments, regardless of aperture, as seen in \autoref{fig:evo}.  The difference between the EQG and non-EQG populations begins right around the same time as the differences in SHMR and $\mbh$, suggesting a causal correlation \citep[see][for more discussions]{Cui2021}. Hence it appears that environment, and particularly early environment, is a key factor in determining whether a galaxy will be an EQG.

Figure~\ref{fig:localmassdens}  explores this connection to environment more quantitatively.  This shows the normalised histogram of galaxies for three samples: EQGs (red solid lines), cluster B non-EQGs (blue dashed lines), and cluster A (green dot-dashed lines).  We show this at two representative redshifts: the second row shows the time of EQG selection at $z=3$, while first row shows their progenitors in an earlier epoch at $z=5$.  We vary the aperture over $0.3, 1, 3$~cMpc (left to right panels) to test sensitivity to this value, but the general trends are similar regardless.

Cluster A galaxies (green lines) live in less dense environments than cluster B galaxies, either EQG or non-EQG.  This is expected because cluster A galaxies tend to be lower mass, and hence reside smaller density peaks in the mass distribution.  Hence as mentioned earlier we find a clear mass-density relation.  This is true both at $z=3$ and $z=5$.

Comparing the EQGs (red) and non-EQGs (blue) illustrates our earlier qualitative statement that EQGs tend to live in less dense environments at $z=3$, as shown by the red solid line being shifted to lower overdensities relative to the blue solid line.  However, at $z=5$, this ordering is reversed, and the EQG progenitors actually live in more overdense regions than their comparable-mass $z=3$ progenitors.

We can thus surmise that it is the denser early environment of EQGs that is driving the accelerated growth, and thereby the early quenching.  What is interesting, however, is that this environmental difference mostly disappears by the time of quenching.  The simplistic (linear) view of structure formation is that all density perturbations grow similarly, so early denser environments will always stay denser.  But in actuality this is not always true, and EQGs seem to arise from a special case where an early dense environment allows rapid growth, but as it transitions to a lower density regime, this enables the AGN feedback to be more effective in removing gas and thus quenching the galaxy.

\subsection{Rejuvenating vs. non-rejuvenating EQGs}

\begin{figure}
    \includegraphics[width=0.95\linewidth]{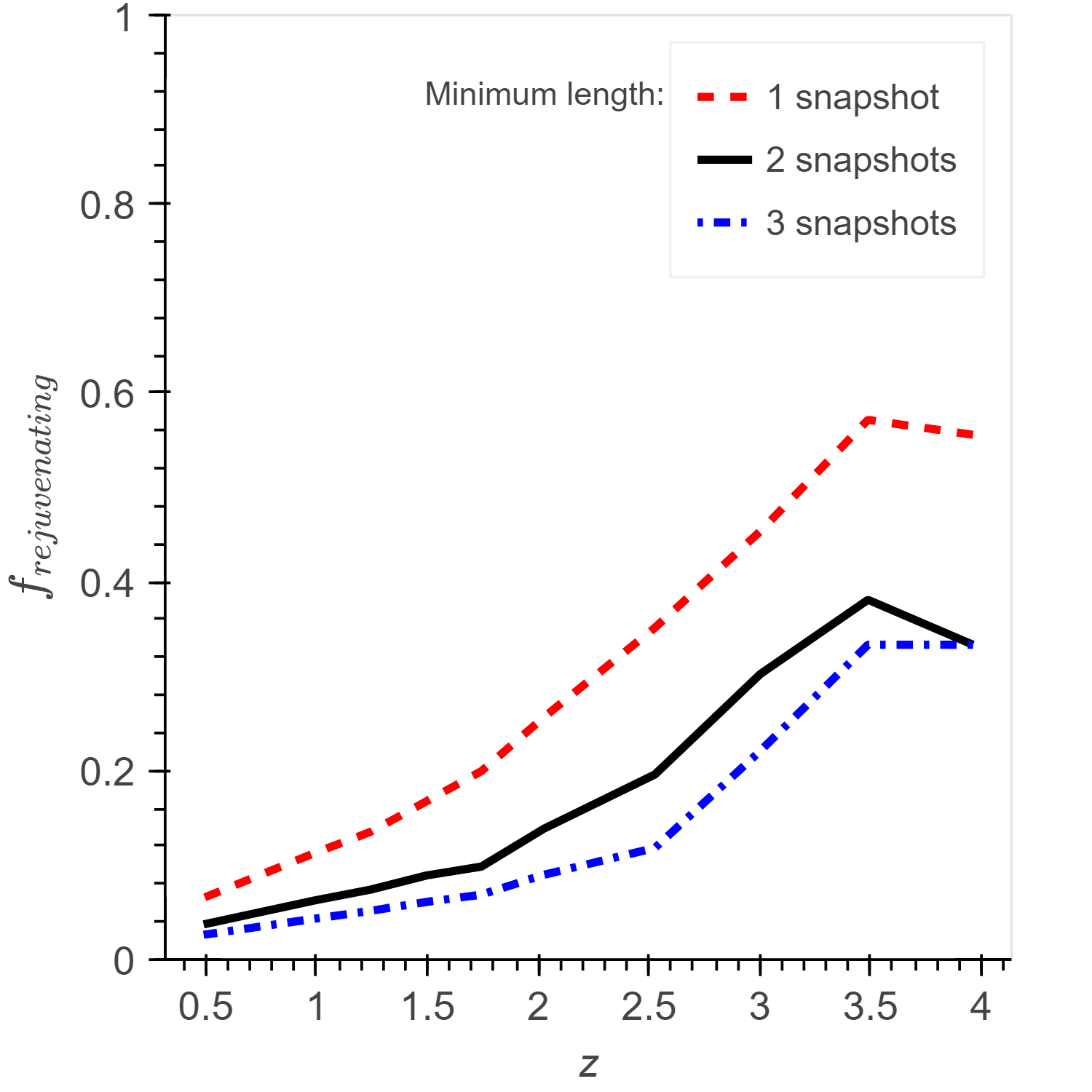}
    \caption{Fraction of quiescent galaxies which rejuvenate after initially quenching as a function of redshift. Lines correspond to the number of consecutive snapshots a galaxy must remain above the rejuvenation threshold to be considered rejuvenated. There is high sensitivity related to the minimum required duration of rejuvenation, with the fraction varying by a factor of $2\times$. The overall trend remains the same, With rejuvenation being much more likely at higher redshifts.}
    \label{fig:rej}
\end{figure}

\begin{figure*}
    \includegraphics[width=0.95\linewidth]{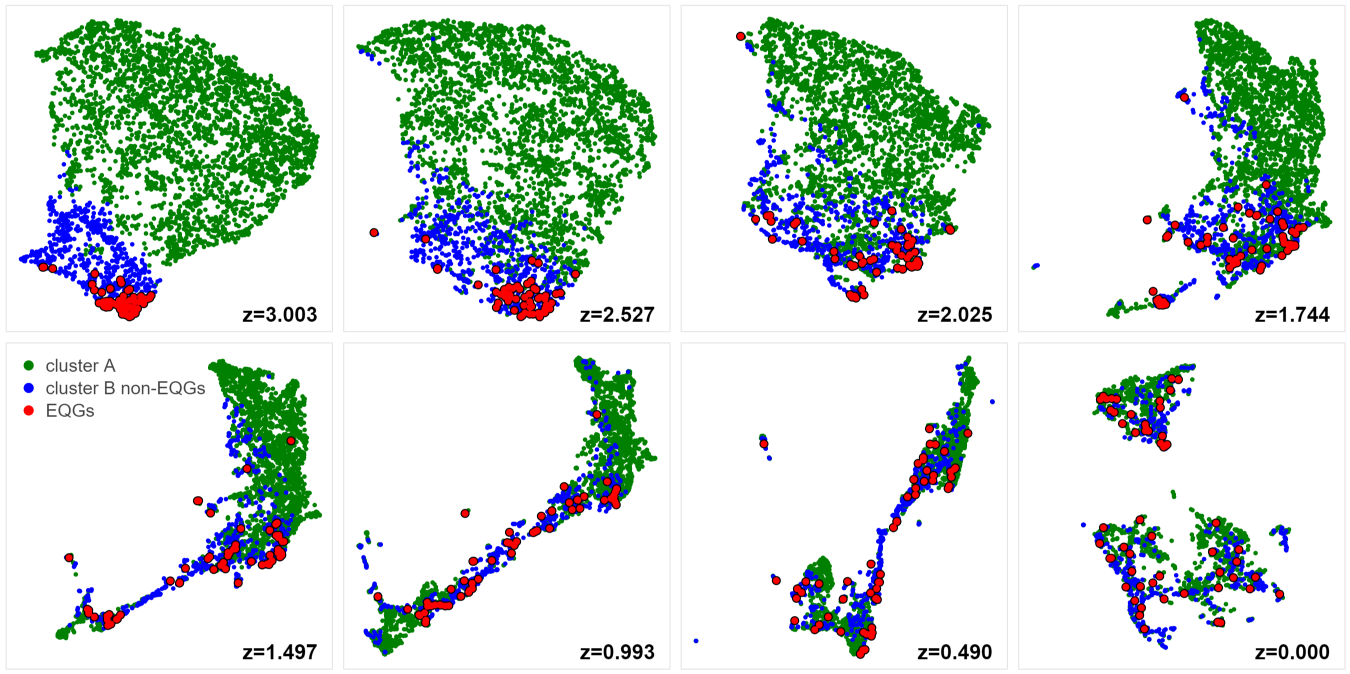}
    \caption{Selected AlignedUMAP embeddings of descendants of $z=3$ massive central galaxies ($\mstellar>10^{9.5}\msolar$). Models were trained on all snapshots from $z=9.642$ to $z=0$. Data-points are colored based on the clustering as described in \S\ref{sec:UMAP}, with descendants of cluster A galaxies in green, EQGs in red and non-EQG cluster B galaxies in blue. We observe a gradual loss of coherence of the region comprised by descendants of EQGs: by $z=2$ they spread throughout the region composed of other cluster B galaxies; by $z=1.5$ there is significant mixing between descendants of cluster A and cluster B galaxies with formation of new structure; by $z=0$ galaxies form two distinct clusters with no relation to the initial groups. This strongly suggests that early quenching is quickly "forgotten", with little categorical difference between descendants of cluster A, cluster B non-EQGs and EQGs by $z=1.5$.}
    \label{fig:umap_progen}
\end{figure*}

So far we have considered the differences between non-EQGs and EQGs, where we did not distinguish EQGs based on whether they rejuvenate mainly because this seemed to make little difference in their global evolution.  However, in detail there are some small differences that may point towards the cause for why some EQGs rejuvenate and others don't.

The red and green lines in Figure~\ref{fig:evo} show the global mean evolution of non-rejuvenating and rejuvenating EQGs, respectively.  The differences between these are much less obvious relative to the non-EQGs.  Particularly, the earlier evolution of these two samples seem to be essentially exactly identical, at least at $z=2,3$.  At $z=4$ there are more significant differences, but the small sample (9 EQGs) make it difficult to assess the robustness of those trends.

Remarkably, there is not even much difference in the post-selection sSFR; there is only a hint at $z=3,4$ that rejuvenating galaxies have higher sSFR.  This is somewhat surprising because rejuvenation by definition implies a higher sSFR.  In point of fact, it appears that the sSFR of the combined sample of rejuvenating galaxies is not particularly higher, and that individual galaxies rejuvenate at various times but never dominate the overall population.  There is some hint at $z=3,4$ that the sSFR is slightly higher immediately after the selection time, but this difference disappears by $z=2$.

The black hole masses are also quite similar, although at $z=2$ there is a hint that non-rejuvenating galaxies have slightly larger BHs.  This may provide somewhat more energy deposition that is able to quench galaxies more permanently.  Similarly, there is not much difference in the SHMR, although again there is a hint at $z=2,4$ that non-rejuvenaters have slightly higher SHMR.  

Environment also does not show large differences, but rejuvenating EQGs seem to live in slightly denser environments. Going back to the examples in Figure~\ref{fig:typical}, the denser environment could contribute to an enhanced merger rate that, at least in that one example, seems to correlate with both quenching (by growing a large BH) and subsequently rejuvenation (from possibly merging with a gas-rich galaxy).  The AGN feedback energy input is also not very different, although there is a hint that at least at $z=4$ the rejuvenating EQGs have a higher cumulative energy input. 

The lack of differences in global properties of rejuvenating vs. non-rejuvenating EQGs suggests that it is some stochastic process that determines whether a galaxy rejuvenates.  The most likely candidate is mergers which brings in fresh fuel to re-ignite star formation temporarily, as hinted by the example shown in Fig.~\ref{fig:typical}.  However, we have also found many cases where there is no obvious merger associated with rejuvenation, so it could also be a rapid set of minor mergers or infalling gas clouds. At lower redshifts stellar mass loss and cooling flows have been postulated to spark rejuvenation, but it is unclear these would have sufficient time to operate in the early universe.

\subsection{Rejuvenation fractions}
\label{sec:rejuvenation}

Figure \ref{fig:rej} shows the fraction of quiescent galaxies that rejuvenate at any point after being selected as quenched at a given redshift.  We previously defined rejuvenation as a period of at least two consecutive snapshots when a previously quenched galaxy has $\text{sSFR}>\frac{1}{\tau_H}$ and present fractions for different minimum required length, which we show as the black line.  The red and blue lines show the result of changing this definition to just 1 (red) or 3 (blue) consecutive snapshots.  For reference, the time between snapshots close to $z=4,3,2,1$ is $48,60,81,120$ Myr.

There is some sensitivity related to this definition; including shorter rejuvenations increases the fraction by a factor of $\ga 1.5\times$. Increasing the minimum required duration of rejuvenation decreases the rejuvenating fraction at all redshifts by a similar factor.  Thus one must be careful in comparing among predictions of rejuvenating fractions, or to observations.  Unfortunately given our snapshot cadence we cannot be more precise in our quantification, but nonetheless we can identify some general trends that are likely to hold regardless of the specific definition.

A clear trend is that rejuvenations at higher redshifts are much more common and very frequent with $f_{\text{rejuvenating}}>0.3$ at $z\geq 2$, falling quickly at lower redshifts to less than 5\% of $z=0.5$ galaxies experiencing rejuvenation by $z=0$.  One factor may be that there is less time to $z=0$ from lower redshifts which could potentially lower $f_{\text{rejuvenating}}$, but in actuality rejuvenation typically happens fairly quickly after quenching if it happens.  Hence the dropping rejuvenation fraction is probably a physical effect, related to the fact that at higher redshifts gas is denser with more gas-rich mergers which is able to resuscitate star formation in galaxies more often.

\subsection{\texorpdfstring{$z=0$ descendants of EQGs}{z=0 descendants of EQGs}}
An interesting question is whether EQGs are systematically different from other quenched galaxies by $z=0$, by virtue of their early quenching. We explore this question by constructing a new set of AlignedUMAP models for the evolving dataset of EQGs and similar massive galaxies. In AlignedUMAP, models for a series of datasets for a given set of objects are trained together, in this case the time series of the progenitors and descendants of $z=3$ massive central galaxies ($\mstellar>10^{9.5}\msolar$) -- the same population as discussed in \S\ref{sec:UMAP} -- from $z=9.642$ to $z=0$. We use $\tt{alignment\_window\_size}$ and $\tt{alignment\_regularisation=0.01}$ as AlignedUMAP parameters, leaving all other parameters unchanged to \S\ref{sec:UMAP}. By $z=0$ many galaxies have merged, so the total number of descendants is fewer than the selection population at $z=3$.

Figure \ref{fig:umap_progen} shows selected AlignedUMAP embeddings from $z=3$ to $z=0$, with colors corresponding to previously defined groups: descendants of cluster A (green), cluster B non-EQGs (blue) and EQGs (red). The overall morphology of the UMAP is different than before because the training set has changed, but it is still the case that at $z=3$, the three groups still occupy very distinct regions in the UMAP space. 

Evolving forward in time, we see a steady decoherence of the three groups identified at $z=3$.  By $z=2.5$ a few EQGs begin to spread across the region populated by non-EQG cluster B galaxies; at $z=2$ there is even some mixing between cluster A and cluster B galaxies.  From $z=2\to 1$, the descendants of EQGs are increasingly mixed in the general population and fail to form an identifiable distinct region within the embedding.  From $z=1\to 0$, we see the emergence of a new (lower) cluster in the UMAP, separating off from the main upper cluster; this turns out to host galaxies that are quenched (at that epoch). By $z=0$ there are now two distinct UMAP clusters, with descendants of the original cluster A, B and EQGs spread roughly evenly across them.

EQGs thus become less distinct in UMAP space and start to spread  throughout the embedding as the number of quenched galaxies increases with time.  This suggests that EQGs are not particularly special even among quiescent galaxies, only among the first to go through a similar quenching process. 

The two new clusters at $z=0$ are strongly identified by the presence of gas and dust within the galaxy. The bottom cluster galaxies have no gas or dust and tend to be fully quenched, while the upper cluster contains galaxies with relatively high gas fractions $\fgas>10^{-2}$ and some still star-forming galaxies. Interestingly, many EQGs appear within the upper $z=0$ cluster, indicating they have substantial cold gas, dust, and sometimes star formation.

In sum, while EQGs are strongly distinct in sSFR at $z=3$, we observe that categorical differences between their descendants and those of other massive galaxies disappear fairly quickly over time.  EQGs do not end up as the most massive galaxies and/or with the largest black holes by $z=0$, and indeed some can even rejuvenate and end up star-forming by $z=0$.  

\section{Conclusion} \label{sec:conclusion}

In this paper, we analysed the properties and evolution of early quenched galaxies in a new (100$\hmpc)^3$, $2\times 1024^3$-particle run of the \simbac simulation~\citep{Hough2023}. This model tweaks $\simba$ in several important ways, yielding improved agreement with key observed demographics.  We identify EQGs as those lying $4\sigma$ below the star-forming main sequence at selected redshifts.  We compare this definition to an observationally-motivated definition based on galaxy photometry from \citet{Gould2023} and find that the resulting statistics are very similar, and at $z=3$ these correspond roughly to a sSFR threshold of $10^{-10}$yr$^{-1}$. Using this definition, we find substantial numbers of EQGs in \simbac as early as $z=5.02$.  In this work we explore the nature of these EQGs, how they differ in properties relative to galaxies that are otherwise similar but not quenched, and the history and fate of EQGs.  Our goal is to combine these investigations into a coherent story for the origin and nature of EQGs in the \simbac simulation.

Our findings can be summarised as follows:
\begin{enumerate}
    \item \simbac produces a reasonable number of early quenched galaxies:  1\% of all galaxies at $z=4$, 5\% at $z=3$ and 17\% at $z=2$, the majority with $10^{-10}<\text{sSFR}<10^{-11}yr^{-1}$ and some highly quenched $\text{sSFR}<10^{-11}yr^{-1}$.  When using a photometric selection to do an apples-to-apples comparison with observations, \simbac broadly agrees with observations at $z\la 4$, but slightly under-predict the number density at $z\ga 4$ by a factor of $2\times$.  This may be a genuine failing of the model, or else a manifestation of the finite volume which does not include the rarest objects.
    \item Comparing to observations broken down by stellar mass, \simbac predicts that the first early quenched galaxies are not necessarily the most massive at that epoch, and there is a hint in \simbac that the quenched fraction actually drops with $M_\star$ at the highest masses which is also seen in observations.  The fraction of quiescent galaxies grows with time steadily in all stellar mass bins, accelerating for $\mstellar>10^{11}\msolar$ for $z<2.5$. 
    
    \item We utilise UMAP to visually evaluate the distinguishing properties of EQGs in \simbac. We find an increased black hole to stellar mass ratio to be one of the key features of EQGs, with all EQGs having $\frac{\rm M_{\rm BH}}{M_\star}>10^{-3}$ and the median 0.3dex higher than other similarly massive galaxies. They also have less rotational support, less dust, lower Eddington ratios, and live in slightly less dense environments.  In contrast, relative to other massive non-EQGs, they have a similar $M_\star$ and halo gas temperature, indicating that these features are not defining characteristics of EQGs.  As noted before, EQGs are not the highest $\mstellar$ galaxies, instead the first EQGs have $M_\star\approx10^{10.5}\msolar$. We argue that the morphological and dust differences are likely to be consequences of early quenching, while the higher black hole fractions and lowered Eddington ratios that typically lie below the threshold for AGN jet and X-ray feedback are the drivers of early quenching.
    
    \item Examining the evolution of individual EQGs selected at $z=3$, we see a consistent pattern that black holes grow quickly until they reach $\sim 10^8M_\odot$ at $z\sim 4.5$.  This lowers the Eddington ratios sufficiently to result in the onset of AGN jet feedback, which together with associated X-ray feedback drives rapid quenching within a couple of hundred Myr.  The AGN power peaks as high as $10^{46}$~erg~s$^{-1}$, and the cumulative AGN feedback energy is typically $\sim 10^{59-60}$~erg for EQGs. These trends holds for both galaxies that quench permanently, and those than later rejuvenate.  The onset of jet and X-ray feedback just prior to quenching occurs in virtually every EQG in \simbac.  The lack of an increase in the CGM temperature points towards X-ray feedback, which acts locally and spherically but with a modest energy input, being key for quenching.
    
    \item Examining the evolution statistically relative to the non-EQG population, we find that the evolution of EQGs diverges early on ($z\ga 6$) from otherwise comparable galaxies.  At $z>4$, EQGs progenitors have higher black hole masses, higher stellar-to-halo mass ratios, emit more AGN kinetic power, and live in denser environments. Curiously, by the actual time of selection, all these quantities become much closer to or inverted relative to the non-EQG population.

    \item Rejuvenating EQGs are not strongly distinct in their evolution of global properties relative to non-rejuvenating EQGs.  However there is a hint particularly at $z\ga 3$ that rejuvenating systems live in slightly denser environments versus non-rejuvenating systems.  Our example rejuvenating galaxy had several jumps in black hole mass suggesting mergers both near the time of quenching and near the time of rejuvenation, but there are many examples rejuvenations that do not show such merger signatures.  Nonetheless, it appears that accretion of a significant gas reservoir in some form is crucial for rejuvenation; it is unlikely to be a smooth process.

    \item Rejuvenation is quite common at early epochs, with $\ga 30\%$ of all quenched galaxies at $z\geq 3$ eventually rejuvenating for at least some period of time. The rejuvenation fractions drop precipitously towards low redshifts, reaching $\sim 5\%$ at $z=0.5$. The exact number is sensitive to the definition of rejuvenation, cautioning comparisons among various models or versus observations. The trend with redshift is expected under the idea that galaxies in the early Universe are subject to more interactions, mergers, and clumpy accretion.
    
    \item An AlignedUMAP analysis of descendants of $z=3$ massive galaxies reveals that the categorical differences are quickly lost. By $z=2$ descendants of EQGs are not obviously distinguishable from descendants of similar massive galaxies, and by $z=1.5$ the initial structure is completely lost, with significant mixing between descendants of cluster A, cluster B non-EQGs and EQGs. This indicates that the memory of early quenching largely disappears by low redshifts, and in fact some $z=3$ EQGs end up with substantial cold dust, gas, and star formation at $z=0$.
    
\end{enumerate}

Our results highlight an overall picture where EQGs occur in special environments that are overdense relative to that of similar-mass systems at early epochs, but end up being comparably dense by the time of quenching.  In such cases, the high early overdensity fuels rapid black hole growth, leading to an earlier onset of AGN jet feedback due to lower Eddington ratios.  However, at the time this turns on, the environment is no longer strongly overdense, which allows the gas to be evacuated more easily, resulting in rapid quenching. We expect that this occurs due to a combination of the AGN jet feedback which deposits energy into the immediate vicinity of the galaxy, as well as X-ray feedback which turns on when the gas fractions are low and drives out or destroys any remaining cold gas from the system. We find that the AGN feedback power consistently peaks at $\sim 10^{45}$~erg~s$^{-1}$, and the cumulative AGN energy from EQGs is $\sim 10^{59\pm 1}$~erg. The combination of these processes results in the destruction of most of the dust in the galaxy, as well as suppressing further accretion which would sustain a rotationally-supported disk leading to more elliptical-like morphologies.  A significant fraction of these EQGs rejuvenate, though that fraction drops quickly to lower redshift, and by $z=0$ the galaxies that were EQGs are not obviously distinguishable from similarly massive early non-quenched galaxies.

Our results suggest that the observations of early quenched galaxies are not a strong challenge to modern galaxy formation models, at least in their current numbers.  This is in agreement with other studies of high-$z$ quenched galaxies, such as \citet{Remus2023} in the Magneticum simulation, who likewise found similar levels of rejuvenation as that seen in \simbac.  Our predictions are limited at the highest redshifts by \simbac's simulation volume, since very early quenching specifically occurs in the rarest regions that are strongly overdense at the earliest epochs.  As observations continue to accumulate on the properties of EQGs, this will provide further tests of the model. Nonetheless, our initial analysis of EQGs here suggests that \simbac is a viable platform to elucidate the origin and evolution of EQGs.

\section*{Acknowledgements}

DR is supported by the Simons Foundation. This work is supported by STFC AGP Grant ST/V000594/1.  WC is also supported by the Atracci\'{o}n de Talento Contract no. 2020-T1/TIC-19882 granted by the Comunidad de Madrid in Spain and the science research grants from the China Manned Space Project. He also thanks the Ministerio de Ciencia e Innovación (Spain) for financial support under Project grant PID2021-122603NB-C21 and ERC: HORIZON-TMA-MSCA-SE for supporting the LACEGAL-III project with grant number 101086388.  For the purpose of open access, the author has applied a Creative Commons Attribution (CC BY) licence to the Author Accepted Manuscript version arising from this submission.

\section*{Data Availability}

The \simbac\ simulation is publicly available through the Flatiron Institute portal at {\tt https://binder.flatironinstitute.org/}; contact the corresponding authors for the Owner and Project number to enable access. The \simbac analysis codes underlying this article will be shared on reasonable request to the corresponding author.

\bibliographystyle{mnras}
\bibliography{refs}

\section*{Appendix A: UMAP input and parameters}
All UMAP models presented in this work are trained on the same set of 62 quantities available in $\caesar$: \texttt{ngas,  nstar,  nbh,  sfr,  sfr\_100,  bhmdot,  bh\_fedd,  masses: [gas, stellar,  dust, bh, HI, H2],  radii: [gas\_20, gas\_hm, gas\_80,  stellar\_20, stellar\_hm, stellar\_80, bh\_20, bh\_hm, bh\_80, baryon\_20, baryon\_hm,  baryon\_80, total\_20, total\_hm,  total\_80], metallicities: [stellar],  velocity\_dispersions: [gas, stellar, bh, baryon, total],  rotation: [gas\_kappa\_rot, stellar\_kappa\_rot, baryon\_kappa\_rot, total\_kappa\_rot], ages: [mass\_weighted, metal\_weighted],  local\_mass\_density: [300, 1000, 3000], local\_number\_density: [300, 1000, 3000],  masses (halo): [gas, stellar, dust, bh, HI, H2, total], virial\_quantities (halo): [circular\_velocity, spin\_param,  temperature, m500c, r500c]}. For their full description see the $\caesar$ documentation.

\begin{figure*}
    \includegraphics[width=0.83\linewidth]{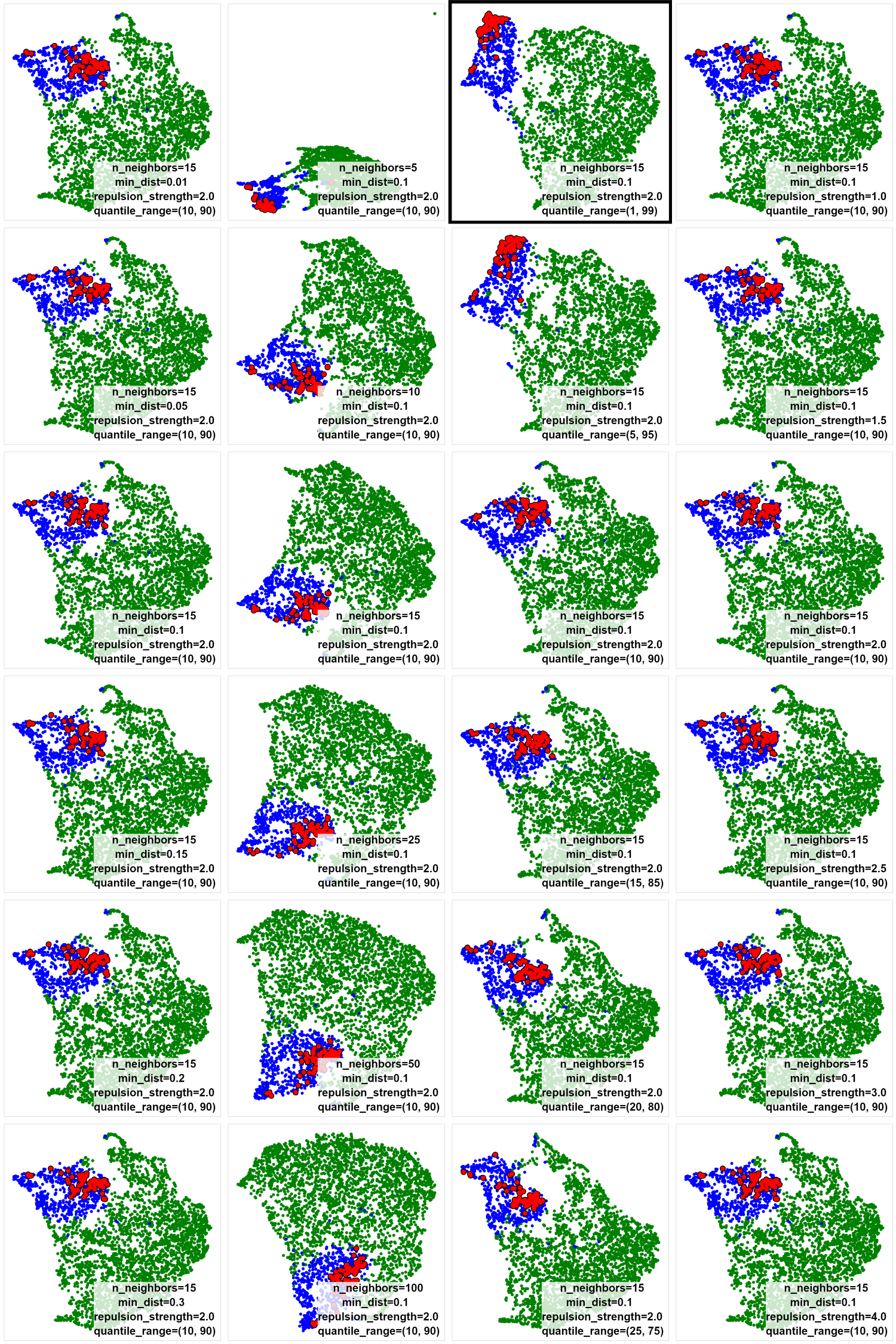}
    \caption{AlignedUMAP exploration of the impact of parameters on the UMAP embedding shape and substructure for $z=3$ massive galaxies. Data-points are colored as defined in \S\ref{sec:UMAP}, with cluster A galaxies in green, EQGs in red and cluster B non-EQGs in blue. The selected parameters are highlighted with a black border, chosen for the spatial separation between clusters A and B as well as low EQG spread within cluster B.}
    \label{fig:umap_param}
\end{figure*}

For completeness we present a simple AlignedUMAP parameter search as Figure \ref{fig:umap_param}. We train 4 sets of models (columns) with ${\tt alignment\_window\_size=1}$ and ${\tt alignment\_regularisation=10^{-3}}$ to evaluate the effects of varying the most impactful parameters: ${\tt n\_neighbors}$, ${\tt min\_dist}$, ${\tt repulsion\_strength}$ and the RobustScaler ${\tt quantile\_range}$. The parameters of each model are shown on each panel. The embeddings are colored by the groups as identified in \S\ref{sec:UMAP}, with cluster A galaxies in green, EQGs in red and cluster B non-EQGs in blue. The embedding highlighted with a black border shows the final selection of parameters, chosen for the clearest visual distinction of cluster A and cluster B as well as the tight grouping of EQGs. The most significant parameters for the shape and substructure of the UMAP embeddings are ${\tt n\_neighbors}$ and ${\tt quantile\_range}$ as expected, with cluster B disappearing for ${\tt n\_neighbors}\geq100$. As the two cluster structure with EQGs isolated to a single region is very consistent across our explored parameter space, we conclude the embedding is stable and reliable.

\label{lastpage}

\end{document}